\documentclass[12pt,dvips]{article}
\usepackage[a4paper]{geometry}
\usepackage{supertabular,lscape,epsfig}
\usepackage{amssymb}
\usepackage{amsmath}
\DeclareSymbolFont{ppa}{OT1}{ppl}{m}{it}
\DeclareMathSymbol{\vv}{\mathalpha}{ppa}{'166}

\begin{document}

\newcommand{\dd}{\,{\rm d}}
\newcommand{\ie}{{\it i.e.},\,}
\newcommand{\etal}{{\it et al.\ }}
\newcommand{\eg}{{\it e.g.},\,}
\newcommand{\cf}{{\it cf.\ }}
\newcommand{\vs}{{\it vs.\ }}
\newcommand{\zdot}{\makebox[0pt][l]{.}}
\newcommand{\up}[1]{\ifmmode^{\rm #1}\else$^{\rm #1}$\fi}
\newcommand{\dn}[1]{\ifmmode_{\rm #1}\else$_{\rm #1}$\fi}
\newcommand{\upd}{\up{d}}
\newcommand{\uph}{\up{h}}
\newcommand{\upm}{\up{m}}
\newcommand{\ups}{\up{s}}
\newcommand{\arcd}{\ifmmode^{\circ}\else$^{\circ}$\fi}
\newcommand{\arcm}{\ifmmode{'}\else$'$\fi}
\newcommand{\arcs}{\ifmmode{''}\else$''$\fi}
\newcommand{\MS}{{\rm M}\ifmmode_{\odot}\else$_{\odot}$\fi}
\newcommand{\RS}{{\rm R}\ifmmode_{\odot}\else$_{\odot}$\fi}
\newcommand{\LS}{{\rm L}\ifmmode_{\odot}\else$_{\odot}$\fi}

\newcommand{\Abstract}[2]{{\footnotesize\begin{center}ABSTRACT\end{center}
\vspace{1mm}\par#1\par
\noindent
{~}{\bf Key words: }{\it #2}}}

\newcommand{\TabCap}[2]{\begin{center}\parbox[t]{#1}{\begin{center}
  \small {\spaceskip 2pt plus 1pt minus 1pt T a b l e}
  \refstepcounter{table}\thetable \\[2mm]
  \footnotesize #2 \end{center}}\end{center}}

\newcommand{\TableSep}[2]{\begin{table}[p]\vspace{#1}
\TabCap{#2}\end{table}}

\newcommand{\FigCap}[1]{\footnotesize\par\noindent Fig.\  %
  \refstepcounter{figure}\thefigure. #1\par}

\newcommand{\TableFont}{\footnotesize}
\newcommand{\TableFontIt}{\ttit}
\newcommand{\SetTableFont}[1]{\renewcommand{\TableFont}{#1}}

\newcommand{\MakeTable}[4]{\begin{table}[htb]\TabCap{#2}{#3}
  \begin{center} \TableFont \begin{tabular}{#1} #4 
  \end{tabular}\end{center}\end{table}}

\newcommand{\MakeTableSep}[4]{\begin{table}[p]\TabCap{#2}{#3}
  \begin{center} \TableFont \begin{tabular}{#1} #4 
  \end{tabular}\end{center}\end{table}}

\newcommand{\Section}[1]{\begin{center}{\section{{\normalsize{#1}}}}\end{center}}
\renewcommand{\thesection}{{{\normalsize{\arabic{section}.}}}}
\newcommand{\Subsection}[1]{\subsection{#1}}
\newcommand{\Acknow}[1]{\par\vspace{5mm}{\bf Acknowledgements.} #1}

\newenvironment{references}%
{
\footnotesize \frenchspacing
\renewcommand{\thesection}{}
\renewcommand{\in}{{\rm in }}
\renewcommand{\AA}{Astron.\ Astrophys.}
\newcommand{\AAS}{Astron.~Astrophys.~Suppl.~Ser.}
\newcommand{\ApJ}{Astrophys.\ J.}
\newcommand{\ApJS}{Astrophys.\ J.~Suppl.~Ser.}
\newcommand{\ApJL}{Astrophys.\ J.~Letters}
\newcommand{\AJ}{Astron.\ J.}
\newcommand{\IBVS}{IBVS}
\newcommand{\PASP}{P.A.S.P.}
\newcommand{\Acta}{Acta Astron.}
\newcommand{\MNRAS}{MNRAS}
\renewcommand{\and}{{\rm and }}
\section{\begin{center}{\small{\rm REFERENCES}}\end{center}}
\sloppy \hyphenpenalty10000
\begin{list}{}{\leftmargin1cm\listparindent-1cm
\itemindent\listparindent\parsep0pt\itemsep0pt}}%
{\end{list}\vspace{2mm}}

\def\TYLDA{~}
\newlength{\DW}
\settowidth{\DW}{0}
\newcommand{\dw}{\hspace{\DW}}

\newcommand{\refitem}[5]{\item[]{#1} #2%
\def\REFARG{#3}\ifx\REFARG\TYLDA\else, {\it#3}\fi
\def\REFARG{#4}\ifx\REFARG\TYLDA\else, {\bf#4}\fi
\def\REFARG{#5}\ifx\REFARG\TYLDA\else, {#5}\fi.}

\pagestyle{myheadings}
\thispagestyle{empty}

                                                                                                         
\newfont{\bb}{ptmbi8t at 12pt}
\newcommand{\xrule}{\rule{0pt}{2.5ex}}
\newcommand{\xxrule}{\rule[-1.8ex]{0pt}{4.5ex}}
\def\thefootnote{\fnsymbol{footnote}}
\begin{center}
{\Large\bf Binary Lenses in OGLE III EWS Database. Season 2005}
\vskip0.5cm
{\bf by}
\vskip0.5cm
{\bf J.~~ S~k~o~w~r~o~n,
~~M.~~ J~a~r~o~s~z~y~\'n~s~k~i, ~~A.~~ U~d~a~l~s~k~i, ~~M.~~ K~u~b~i~a~k,
~~M.\,K.~~ S~z~y~m~a~\'n~s~k~i, ~~G.~~ P~i~e~t~r~z~y~\'n~s~k~i, ~~I.~~
S~o~s~z~y~\'n~s~k~i, ~~O.~~ S~z~e~w~c~z~y~k, ~~\L.~~ W~y~r~z~y~k~o~w~s~k~i
~~and~~ K.~~ U~l~a~c~z~y~k}
\vskip3mm
{Warsaw University Observatory,
Al.~Ujazdowskie~4,~00-478~Warszawa, Poland\\
e-mail:\\
(jskowron,mj,udalski,mk,msz,pietrzyn,soszynsk,szewczyk,wyrzykow,ulaczyk)
@astrouw.edu.pl}
\end{center}
\vskip2cm

\Abstract{We present nine new binary lens candidates from OGLE-III Early
Warning System database for the season of 2005. We have also found four
events interpreted as single mass lensing of double sources. The candidates
have been selected by visual light curves inspection. Examining the models
of binary lenses in our previous studies (10 caustic crossing events of
OGLE-II seasons 1997--1999 and 34 binary lens events of OGLE-III seasons
2002--2004, including one planetary event), in this work and in three
publications concerning planetary events of season 2005, we find four cases
of extreme mass ratio binaries ($q\le0.01$), and almost all other models
with mass ratios in the range $0.1<q<1.0$, which may indicate the division
between planetary systems and binary stars.}{Gravitational lensing --
Galaxy: center -- binaries: general}


\Section{Introduction}
In this paper we present the results of the search for binary lens events
among microlensing phenomena discovered by the Early Warning System (EWS --
Udalski \etal 1994, Udalski 2003) of the third phase of the Optical
Gravitational Lensing Experiment (OGLE-III) in the season of 2005. This is
a continuation of the study of binary lenses in OGLE-II (Jaroszy\'nski
2002, hereafter Paper~I) and OGLE-III databases (Jaroszy\'nski \etal 2004,
hereafter Paper~II and Jaroszy\'nski \etal 2006, hereafter Paper~III). The
results of the similar search for binary lens events in MACHO data were
presented by Alcock \etal (2000).

The motivation of the study remains the same -- we are going to obtain a
uniform sample of binary lens events, selected and modeled with the same
methods for all seasons. The sample may be used to study the population of
binary systems in the Galaxy. The method of observation of the binaries
(gravitational lensing) allows to study their mass ratios distribution,
since they are directly given by the models. The binary separations are
more difficult, because only their projection into the sky expressed in
Einstein radius units enters the models. In small number of cases the
estimation of the masses and distances to the lenses may be possible.

Cases of extremely low binary mass ratios ($q\le0.01$) are usually
considered as planetary lensing. Such events have been discovered in
OGLE-III database for season 2003 (Bond \etal 2004), 2005 (Udalski
\etal 2005, Gould \etal 2006, Beaulieu \etal 2006), and 2007 (to be
published). The adequate modeling of a planetary event requires frequent
round the clock observations of the source, which is achieved by cooperation
of observers at different longitudes on Earth. In cases of less
extreme lenses, the observations of single telescope may be sufficient
to obtain well constrained models of the systems. The present analysis
is based on the OGLE-III data alone.

Our approach follows that of Papers I, II, and III, where the references to
earlier work on the subject are given. Some basic ideas for binary lens
analysis can be found in the review article by Paczy\'nski (1996). Paper~I
presents the analysis of 18 binary lens events found in OGLE-II data with
10 safe caustic crossing cases. There are 15 binary lens events reported in
Paper~II, and 19 in Paper~III.

In Section~2 we describe the selection of binary lens candidates. In
Section~3 we describe the procedure of fitting models to the data. The
results are described in Section~4, and the discussion follows in
Section~5. The extensive graphical material is shown in Appendix.

\Section{Choice of Candidates} 
The OGLE-III data are routinely reduced with difference photometry ({\sc
dia}, Alard and Lupton 1998, Alard 2000) which gives high quality light
curves of variable objects. The EWS system of OGLE-III (Udalski 2003)
automatically picks up candidate objects with microlensing-like
variability.

There are 597 microlensing event candidates selected by EWS in the 2005
season. We visually inspect all candidate light curves looking for features
characteristic for binary lenses (multiple peaks, U-shapes,
asymmetry). Light curves showing excessive noise are omitted. We select 12
candidate binary events in 2005 data for further study. For these candidate
events we apply our standard procedure of finding binary lens models
(\cf Papers~I--III and Section~3).

\Section{Fitting Binary Lens Models}
The models of the two point mass lens were investigated by many authors
(Schneider and Weiss 1986, Mao and DiStefano 1995, DiStefano and Mao 1996,
Dominik 1998, to mention only a few). The effective methods applicable for
extended sources were described by Mao and Loeb (2001). We follow their
approach and use image finding algorithms based on the Newton method. Our
implementation of adaptive contouring approach to image finding (Dominik
2007) proved more time consuming, and has not been used.

We fit binary lens models using the $\chi^2$ minimization method for the 
light curves. It is convenient to model the flux at the time $t_i$ as: 
$$F_i =F(t_i)=A(t_i)\times F_s+F_b\equiv\left(A(t_i)-1\right)\times
F_s+F_0\eqno(1)$$ 
where $F_s$ is the flux of the source being lensed, $F_b$ the blended flux
(from the source close neighbors and possibly the lens), and the
combination $F_b+F_s=F_0$ is the total flux at baseline, measured long
before or long after the event. The last parameter can be reasonably well
estimated with observations performed in seasons preceding and following
2005, as a weighted mean:
$$F_0=\frac{\sum\limits_{i^\prime=1}^{N^\prime}\displaystyle\frac{F_i}{\sigma_i}}
           {\sum\limits_{i^\prime=1}^{N^\prime}\displaystyle\frac{1}{\sigma_i}}\eqno(2)$$
where $F_i$ are the observed fluxes and $\sigma_i$ their estimated
photometric errors. The summation over $i^\prime$ does not include
observations of 2005, and $N^\prime$ is the number of relevant
observations.

In fitting the models we use rescaled errors (compare Papers~I--III). More
detailed analysis (\eg Wyrzykowski 2005) shows that the OGLE photometric
errors are overestimated for very faint sources and underestimated for
bright ones. Error scaling used here, based on the scatter of the source
flux in seasons when it is supposedly invariable, is the simplest
approach. It gives the estimate of the combined effect of the observational
errors and possibly undetectable, low amplitude internal source
variability. We require that constant flux source model fits well the other
seasons data after introducing error scaling factor $s$:
$$\chi_{\rm other}^2=\sum\limits_{i^\prime=1}^{N^\prime}
\frac{(F_i-F_0)^2}{(s\sigma_i)^2}=N^\prime-1.\eqno(3)$$

The lens magnification (amplification) of the source $A(t_i)=A(t_i;p_j)$
depends on the set of model parameters $p_j$. Using this notation one has
for the $\chi^2$:
$$\chi^2=\sum\limits_{i=1}^N\frac{\left((A_i-1)F_s+F_0-F_i\right)^2}
{\sigma_i^2}.\eqno(4)$$
The dependence of $\chi^2$ on the binary lens parameters $p_j$ is
complicated, while the dependence on the source flux is quadratic. The
equation $\partial\chi^2/\partial F_s=0$ can be solved algebraically,
giving $F_s=F_s(p_j;\{F_i\})$, thus effectively reducing the dimension of
parameter space. Any method of minimizing $\chi^2$ may (in some cases) give
unphysical solutions with $F_{\rm s}>F_0$, which would imply a negative
blended flux. To reduce the occurrence of such faulty solutions we add an
extra term to $\chi^2$ which vanishes automatically for physically correct
models with $F_{\rm s}\le F_0$, but is a fast growing function of the
source flux $F_{\rm s}$ whenever it exceeds the base flux $F_0$.

Our analysis of the models, their fit quality etc. is based on the
$\chi_1^2$ calculated with the rescaled errors:
$$\chi_1^2\equiv\frac{\chi^2}{s^2}\eqno(5)$$
which is displayed in the tables and plots. For events with multiple models
(representing different local minima of $\chi^2$), we assess the relevance
of each model with the relative weight $w\sim\exp(-\chi_1^2/2)$.

The number of events with at least one well sampled caustic crossing is
relatively high among 2005 binary lens candidates, so the extended source
models can be fitted. The extended source models may seem more difficult,
but looking for the $\chi^2$ minima may be easier in this case. The light
curves for point sources have (formally) infinite jumps on caustic
crossings. This implies that a small change in model parameters may
drastically change the synthetic light curve shape and the quality of the
fit. For extended sources the model light curves are continuous which
substantially diminishes the problem. This property may be exploited in two
ways. First, postulating a very large source radius one obtains an almost
smooth $\chi^2$ dependence on other parameters, with much lower number of
local minima, which may help in the initial stages of optimization. On the
other hand the optimization of a model with slightly perturbed source size,
may lead to a better fitted new solution. The latter approach is easy to
implement and control and we use it routinely.

The binary system consists of two masses $m_1$ and $m_2$, where by
convention ${m_1\le m_2}$. The Einstein radius of the binary lens is
defined as:
$$r_{\rm E}=\sqrt{\frac{4G(m_1+m_2)}{c^2}\frac{d_{\rm OL}d_{\rm LS}}
{d_{\rm OS}}}\eqno(6)$$
where $G$ is the constant of gravity, $c$ is the speed of light, $d_{\rm
OL}$ is the observer--lens distance, $d_{\rm LS}$ is the lens--source
distance, and $d_{\rm OS}\equiv d_{\rm OL}+d_{\rm LS}$ is the distance
between the observer and the source. The Einstein radius serves as a length
unit and the Einstein time: ${t_{\rm E}=r_{\rm E}/\vv_\perp}$, where
$\vv_\perp$ is the lens velocity relative to the line joining the observer
with the source, serves as a time unit. The passage of the source in the
lens background is defined by seven parameters: ${q\equiv m_1/m_2}$
(${0<q\le1}$) -- the binary mass ratio, $d$ -- binary separation expressed
in $r_{\rm E}$ units, $\beta$ -- the angle between the source trajectory as
projected onto the sky and the projection of the binary axis, $b$ -- the
impact parameter relative to the binary center of mass, $t_0$ -- the time
of closest approach of the source to the binary center of mass, $t_E$ --
the Einstein time, and $r_s$ the source radius. Thus we are left with the
seven or six dimensional parameter space, depending on the presence/absence
of observations covering the caustic crossings.

We begin with a scan of the parameter space using a logarithmic grid of
points in $(q,d)$ plane (${10^{-3}\le q\le1}$, ${0.1\le d\le10}$) and
allowing for continuous variation of the other parameters. The choice of
starting points combines systematic and Monte Carlo searching of regions in
parameter space allowing for caustic crossing or cusp approaching events.
The $\chi^2$ minimization is based on downhill method and uses standard
numerical algorithms. When a local minimum is found we make a small Monte
Carlo jump in the parameter space and repeat the downhill search. In some
cases it allows for finding a different local minimum. If it does not work
several times, we stop and try next starting point.

Some models may be improved by taking into account parallax effects and/or
the changes in the orientation and separation of the binary lens caused by
its rotation. The parallax parameter:
$$\pi_{\rm E}=\frac{1{\rm a.u.}}{{\tilde r}_\mathrm{E}}\qquad
{\tilde r}_{\rm E}=r_{\rm E}\frac{d_{\rm OS}}{d_{\rm LS}}\eqno(7)$$
where ${\tilde r}_{\rm E}$ is the radius of the Einstein ring projected
into the observer's plane, measures the influence of the Earth motion on
the source path relative to the lens position. Another parameter defines
the orientation of the source path relative to the Ecliptic. In the linear
approximation the lens rotation may be described as:
$$d=d_0+{\dot d}(t-t_0)\qquad
\beta=\beta_0+{\dot\beta}(t-t_0)\eqno(8)$$
where the subscript ``0'' denotes values of parameters measured when the
source approaches the binary center of mass. (The detailed definitions of
all binary lens parameters when parallax and rotation are taken into
account are given by Jaroszy\'nski \etal 2005).

Only the events with characteristics of caustic crossing (apparent
discontinuities in observed light curves, U-shapes) can be treated as safe
binary lens cases. The double peak events may result from cusp approaches,
but may also be produced by double sources (\eg Gaudi and Han 2004). In
such cases we also check the double source fit of the event postulating:
$$F(t)=A(u_1(t))\times F_{\rm s1}+A(u_2(t))\times F_{\rm s2}+F_{\rm b}\eqno(9)$$
where $F_{\rm s1}$, $F_{\rm s2}$ are the fluxes of the source components,
$F_{\rm b}$ is the blended flux, and $A(u)$ is the single lens
amplification. In most cases the formula for a point source amplification
(Paczy\'nski 1986) is sufficient, but the influence of the source finite
size may show up, when the amplification is extreme. We routinely take this
effect into account. The dimensionless source--lens separations are given
as:
$$u_1(t)=\sqrt{{b_1}^2+\frac{(t-t_{01})^2}{{t_{\rm E}}^2}}\qquad
u_2(t)=\sqrt{{b_2}^2+\frac{(t-t_{02})^2}{{t_{\rm E}}^2}}\eqno(10)$$
where $t_{01}$, $t_{02}$ are the closest approach times of the source
components, $b_1$, $b_2$ are the respective impact parameters, and $t_{\rm
E}$ is the (common) Einstein time.

\Section{Results}
Our fitting procedures applied to 12 candidate events selected give the
results summarized in Table~1. (We do not include the published models of
planetary events. While for the event OGLE~2005-BLG-071 the modeling based
on OGLE data alone is possible and gives results consistent with the final
model based on combined observations from many telescopes, such an approach
is not possible for OGLE~2005-BLG-169 or OGLE~2005-BLG-390.)
\renewcommand{\TableFont}{\scriptsize}
\MakeTable{c@{\hspace{3pt}}c@{\hspace{3pt}}rcccrrcrcll}{12.7cm}
{Binary lenses parameters, excluding known planetary cases}
{\hline
\noalign{\vskip 5pt}
Event &  & \multicolumn{1}{c}{$\chi_1^2/{\rm DOF}$} & $s$ & 
$q$ & $d$ & \multicolumn{1}{c}{$\beta$} & \multicolumn{1}{c}{$b$} & 
$t_0$ & \multicolumn{1}{c}{$t_E$} & $f$ & $r_s$ & $\pi_{\rm E}$ \\
\noalign{\vskip5pt}
\hline
017 &  d &    902.8/393 &  1.71 & 0.061 & 1.229 &  155.22 &  $-0.14$ &  3456.5 &   57.9 &  0.55 &   &   \\
017 &  d &   1032.6/392 &  1.71 & 0.002 & 1.018 &  141.69 &  $ 0.07$ &  3456.6 &   73.2 &  0.28 &   &  \\
018 &  b &   6190.2/491 &  1.73 & 0.516 & 0.726 &  253.43 &  $ 0.12$ &  3513.0 &   55.5 &  1.00 &  0.0237 &  0.08 \\
062 &  b &    720.1/370 &  1.57 & 0.691 & 2.065 &  128.29 &  $-0.12$ &  3479.6 &   34.7 &  1.00 &  & \\
128 &  b &    467.7/309 &  1.03 & 0.852 & 1.606 &  200.29 &  $-0.03$ &  3511.2 &   50.3 &  0.47 &  & \\
153 &  b &   1395.4/467 &  2.88 & 0.806 & 0.830 &  209.02 &  $ 0.53$ &  3544.5 &   32.1 &  0.96 &  0.0160 &  0.59 \\
189 &  b &   1023.7/771 &  1.04 & 0.790 & 0.842 &  166.02 &  $ 0.75$ &  3526.4 &   98.7 &  0.96 &  0.0011 &  \\
226 &  b &    607.3/353 &  1.13 & 0.284 & 0.296 &   96.86 &  $ 0.02$ &  3573.7 &   37.5 &  0.81 &  0.0088 &  \\
327 &  b &    836.8/441 &  1.09 & 0.810 & 0.566 &  260.05 &  $ 0.14$ &  3566.2 &  112.6 &  1.00 &  0.0008 &  0.19 \\
327 &  b &    885.6/441 &  1.09 & 0.099 & 3.277 &   60.82 &  $ 2.40$ &  4208.9 &  477.3 &  0.97 &  0.0002 &  0.02 \\
331 &  ? &  22752.0/462 &  1.66 & 0.199 & 0.872 &  111.59 &  $ 0.08$ &  3562.6 &   17.8 &  0.69 &  &  \\
463 &  b &    666.4/413 &  1.25 & 0.120 & 1.538 &   83.90 &  $ 0.82$ &  3600.5 &   72.2 &  0.38 &  &  \\
468 &  b &    378.5/261 &  1.21 & 0.529 & 0.440 &   55.23 &  $ 0.00$ &  3592.3 &   52.9 &  0.05 &  0.0005 &  \\
477 &  d &    572.7/403 &  1.32 & 0.009 & 1.374 &  186.97 &  $-0.05$ &  3630.4 &   84.9 &  0.27 &  &  \\
\noalign{\vskip5pt}
\hline
\noalign{\vskip5pt}
\multicolumn{13}{p{14.5cm}}{Note: 
The table contains all 2005 season events, which have been modeled as
binary lenses. The columns show: the event 2005 EWS number, the event
classification (``b'' for binary lens, ``d'' for double source, ``?'' for
unknown), the rescaled $\chi^2_1$, number of DOF,the scaling factor $s$
($\chi^2_1=\chi^2_{\rm raw}/s^2$), the mass ratio $q$, the binary
separation $d$, the source trajectory direction $\beta$, the impact
parameter $b$, the time of the closest center of mass approach $t_0$, the
Einstein time $t_{\rm E}$ and the blending parameter $f\equiv F_{\rm
s}/F_0$. For events with resolved caustic crossings the size of the source
$r_s$ is given; otherwise it is omitted. Models taking into account the
parallax effects have assigned value of the parameter $\pi_{\rm E}$.}}

In the second column of Table~1 we assess the character of the events. In 9
cases (of 12 investigated) the events are safe binary lens phenomena in our
opinion (designated as ``b'' in Table~1). There are 2 cases classified as
double source events (``d'' in Table~1) and 1 event with a low quality fit
(``?'' in Table~1). The source paths and model light curves are shown in
the first part of Appendix.

In modeling of the event OGLE~2005-BLG-018 we have been forced to include
parallax and rotation effects. The light curve of this event shows a smooth
peak near Julian date 2\,453\,460 and two asymmetric tall peaks $\approx50$
days and $\approx70$ days later. The sharp bend of the light curve before
the second tall peak implies that the source is entering the caustic here,
so the tall peaks may be interpreted as two separate enterings into the
caustic region by a large source. The trajectory should also approach a
cusp to obtain the smaller observed peak. Our model fits all the observed
features qualitatively well.

In a few other cases the inclusion of the parallax effect has given models
of substantially better formal quality ($\chi_1^2$ smaller by at least
50). For such models the value of the parallax parameter is included in
Table~1. In the remaining cases the parallax has no significant effect.

The results of double source modeling are summarized in Table~2. The double
source modeling is applied to all binary lens candidates and some other
non-standard events. While formally the fits are usually better for binary
lenses, in two cases we prefer double source models as more natural, giving
less complicated light curves. The comparison of two kinds of fits is given
in the second part of Appendix, and the well separated double source events
-- in the third.

\MakeTable{lcrccccrcc}{12.7cm}{Parameters of double source modeling}
{\hline
\noalign{\vskip5pt}
Event &  & $\chi_1^2/$DOF & $b_1$ & $b_2$ & $t_{01}$ & $t_{02}$ & 
\multicolumn{1}{c}{$t_{\rm E}$} &  $f_1$ & $f_2$ \\
\noalign{\vskip5pt}
\hline
\noalign{\vskip5pt}
 017 & d &    973./393 & 0.0039 & 0.0956 & 3455.75 & 3456.38 &    78.0 & 0.007 & 0.270 \\
 018 & b & 359081./490 & 0.0000 & 0.0000 & 3512.44 & 3528.89 &   130.7 & 0.074 & 0.092 \\
 062 & b &   2741./363 & 0.1999 & 0.0019 & 3460.68 & 3476.40 &    64.1 & 0.162 & 0.030 \\
 066 & d &    419./312 & 0.0095 & 0.0204 & 3439.10 & 3448.92 &   165.0 & 0.015 & 0.024 \\
 128 & b &   5554./310 & 0.1577 & 0.1091 & 3495.45 & 3538.23 &    44.2 & 0.443 & 0.314 \\
 153 & b & 475463./470 & 0.0016 & 0.0000 & 3557.46 & 3560.65 &   313.7 & 0.011 & 0.001 \\
 189 & b &   2945./772 & 0.0264 & 0.0000 & 3502.08 & 3512.75 &   252.8 & 0.068 & 0.002 \\
 192 & d &    687./383 & 1.7234 & 0.0158 & 3516.13 & 3529.50 &    27.1 & 0.994 & 0.006 \\
 226 & b &  26557./354 & 0.0755 & 0.0000 & 3573.31 & 3573.71 &    34.7 & 0.824 & 0.176 \\
 327 & b & 133339./348 & 0.0000 & 0.0000 & 3575.25 & 3579.62 &    74.4 & 0.184 & 0.317 \\
 331 & ? &  10785./433 & 0.0000 & 0.1561 & 3554.27 & 3560.16 &    27.7 & 0.141 & 0.859 \\
 463 & b &  84000./415 & 0.0000 & 0.0000 & 3585.86 & 3609.45 &   312.1 & 0.003 & 0.021 \\
 468 & b &   7596./231 & 0.0230 & 0.0000 & 3589.65 & 3594.56 &    17.0 & 0.126 & 0.004 \\
 477 & d &    521./404 & 0.0599 & 0.0128 & 3629.02 & 3663.12 &    67.8 & 0.336 & 0.033 \\
\noalign{\vskip5pt}
\hline
\noalign{\vskip5pt}
\multicolumn{10}{p{13cm}}{Note: The table contains the event number in
2005 EWS database, the classification of the event, the rescaled $\chi^2$
value and the DOF number, the impact parameters $b_1$ and $b_2$ for the two
source components, times of the closest approaches $t_{01}$ and $t_{02}$,
the Einstein time $t_{\rm E}$, and the blending parameters $f_1=F_{\rm
s1}/(F_{\rm s1}+F_{\rm s2}+F_{\rm b})$ and $f_2=F_{\rm s2}/(F_{\rm
s1}+F_{\rm s2}+F_{\rm b})$.}}

Our sample of binary lenses consists now of $10+15+19+9+3=56$ events of
Papers~I--III, and the present work, some of them with multiple models,
plus three published planetary events of the season 2005, which we also
include. Using the sample we study the distributions of various binary
lens parameters. In Fig.~1 we show the histograms for the mass ratio and the
binary separation. The mass ratio is practically limited to the range $0.1
\le q\le1$ with very small probability of finding a model in the range
$0.01\to0.1$. The four published planetary lens events are also included.
\begin{figure}[htb]
\centerline{
\includegraphics[height=63.0mm,width=62.0mm]{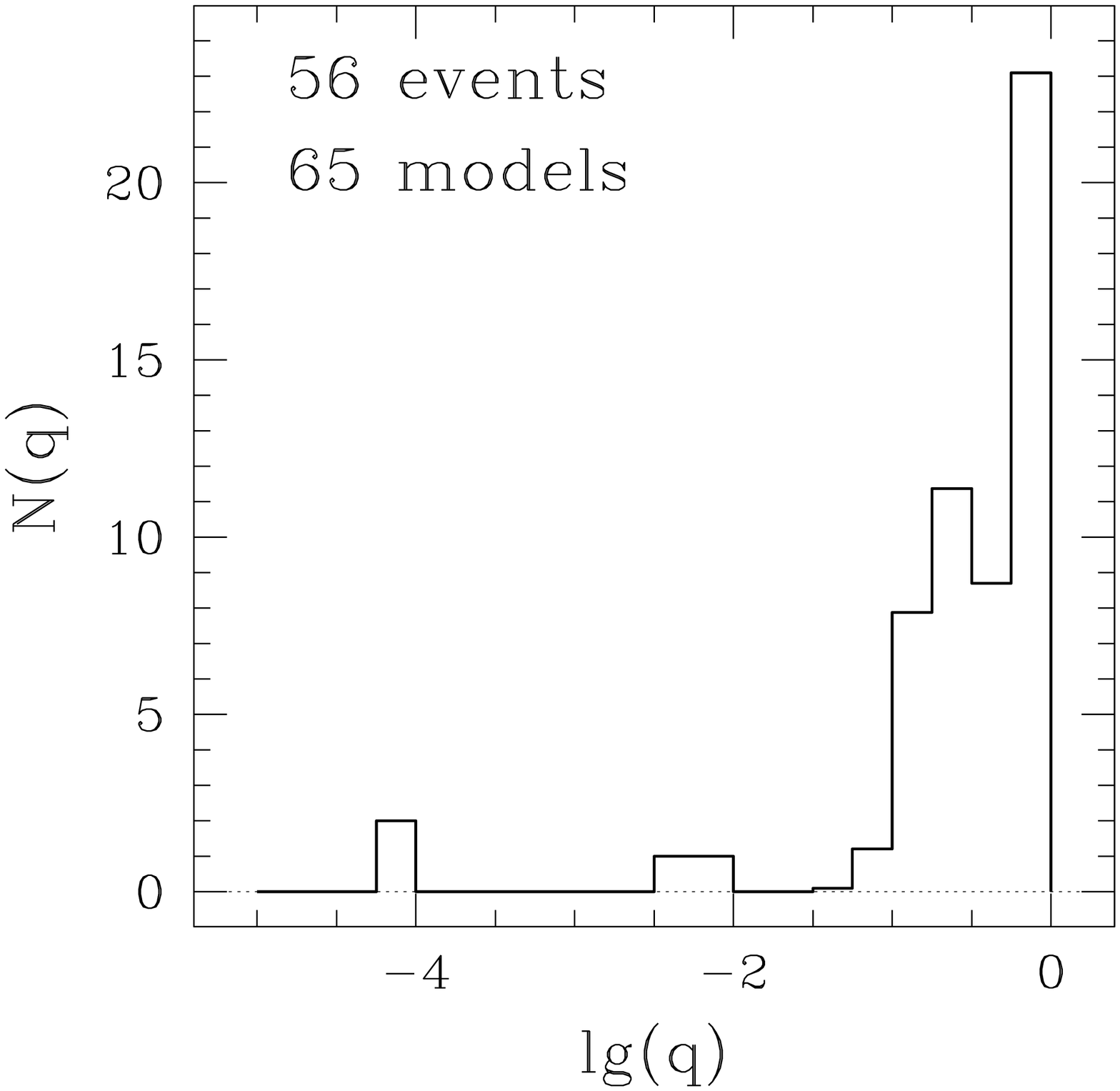} \hfill
\includegraphics[height=63.0mm,width=62.0mm]{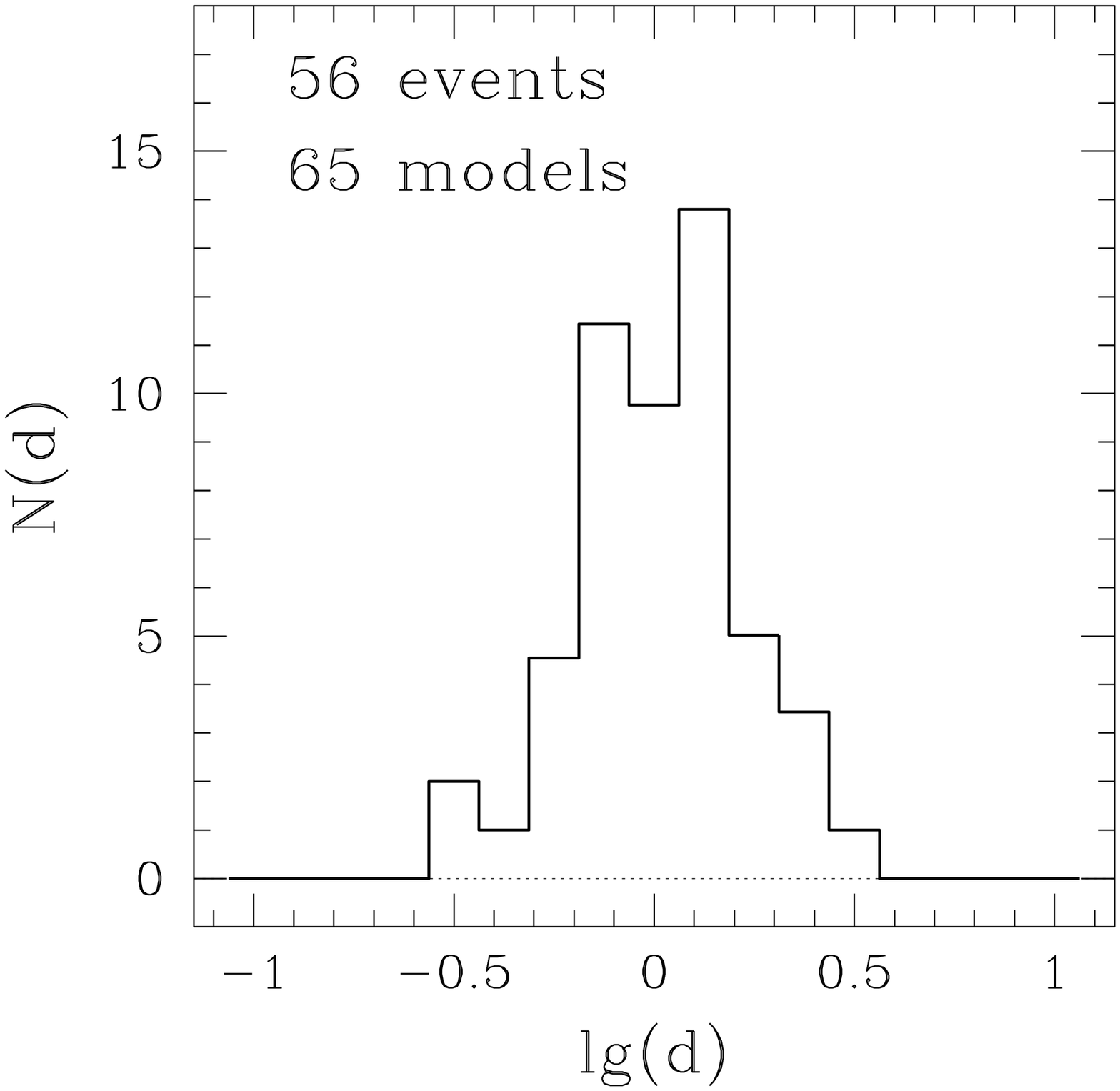}}
\FigCap{The histogram of mass ratios ({\it left}) and separations 
({\it right}) for binary lens events of OGLE-II (Paper~I) and OGLE-III
(Paper II, III, and this work). The histogram includes 56 events, some of
them with multiple models. The alternative models of any event have been
assigned fractional weights.}
\end{figure}

\Section{Discussion}
The caustic crossing events dominate our sample of binary lenses.
According to Night, DiStefano and Schwamb (2007) the number of caustic
crossing and other events caused by binary lenses should be comparable.
This claim is based on simulation investigating binary lenses of various
mass ratios and separations, and source paths with different impact
parameters and directions relative to the binary. The simulated light
curves with some noise added are then checked algorithmically for
differences relative to a point lens light curve, asymmetry, and multiple
peaks. Cases which at the assumed level of ``observational'' noise can be
classified as binary lenses, are then subdivided into smooth and caustic
crossing categories with similar counts. We do not object to this
conclusion, however our aim is to obtain well constrained models of the
lenses, not the list of events which cannot be modeled as point
lenses. The similarity between lensing of a double source and some cases of
binary lensing can be seen in Appendix, and Papers II and III.  Another
example -- OGLE~2005-BLG-055 (Jaroszy\'nski and Paczy\'nski 2002) -- is also
not a point lens case, but its binary lens model is completely
unconstrained. This shows, that events with well marked deviations from
point lens light curves (\eg chosen by visual inspection) have better
chance of getting well constrained models.  Since caustic crossing events
usually imply large deviations from single peak, smooth light curves
characteristic for point lens events, they also outnumber other kinds of
events in our sample.

Our models of the events OGLE~2005-BLG-018 and OGLE~2005-BLG-153 have the
highest $\chi_1^2/{\rm DOF}$ ratio. Simultaneously the model light curves
reproduce all changes of the observed flux quite well. Both sources reach
very high apparent luminosities ($I\approx12$~mag) due to the lens
amplification and in the most interesting time interval are observed many
times during each night. Our method of errors scaling based on the flux
scatter at the baseline is not sufficient in these cases. The application
of Wyrzykowski (2005) error scaling improves the formal quality of the
models substantially, but not completely. The problem may be partially due
to the limb darkening of the source, which we do not include in our models.

Our classification of the investigated events into the binary lens double
source/ unknown categories needs further explanation. The formally best
binary lens model of the event OGLE~2005-BLG-017 does not reflect the
bending of the observed light curve during the nights with Julian dates
3456--3459. Other models which fit this part of the light curve much
better, have unacceptable quality at earlier epochs (compare Appendix and
Table~1). The double source fit is formally better, but the bend of the
light curve is also not well modeled in this case. The double source model
of the event OGLE~2005-BLG-477 is quantitatively better. The best binary
lens model gives a comparable fit to the data but it predicts a caustic
crossing during unobserved period of winter 2005/2006, which we treat as an
unwanted, not verifiable property of the model.

Our sample of OGLE binary lens events contains now 56 cases. The bimodality
of the mass ratio distribution and the lack of intermediate $q$ values
remains a valid interpretation of the data. We are not trying a statistical
interpretation of mass ratio distribution in this paper skipping it into a
further publications.

\Acknow{We thank Shude Mao for the permission of using his binary lens  
modeling software. This work was supported in part by MNiSW grants N203 008
32/0709 and N203 030 32/4275. JS and {\L}W acknowledge support of the
European Community's Sixth Framework Marie Curie Research Training Network
Programme, Contract No. MRTN-CT-2004-505183 ``ANGLES''.}

\vfill
\eject

\begin{center}{\bf Appendix}\end{center}
{\it Binary Lens Models}
\vskip6pt
Below we present the plots for the 12 events for which the binary lens
modeling has been applied. Some of the events, especially cases without
apparent caustic crossing, may have alternative double source models. In
such cases we show the comparison of the binary lens and double source fits
to the data in the next subsection.

The events are ordered and named according to their position in the OGLE
EWS database for the season 2005. We include two models for
OGLE~2005-BLG-017 event, despite the huge difference in their formal fit
quality, to show that it is possible to model the highly amplified part of
the light curve with a binary lens. Alternative models of the event
OGLE~2005-BLG-327 have different caustic topology.

Each case is illustrated with two panels. The most interesting part of the
source trajectory, the binary and its caustic structure are shown in the
left panel for the case considered. The labels give the $q$ and $d$
values. On the right the part of the best fit light curve is compared with
observations. The labels give the rescaled $\chi_1^2$/DOF values. The
source radius (as projected into the lens plane and expressed in Einstein
radius units) is labeled only for the events with resolved caustic
crossings. Below the light curves we show the differences between the
observed and modeled flux in units of rescaled errors. The dotted lines
show the rescaled $\pm3\sigma$ band.

In all cases we plot the source trajectories in the coordinate systems of
the binary lenses. In the case of OGLE~2005-BLG-018 the caustic structure
shown corresponds to the binary separation when source passes the binary
center of mass. The effects of rotation on the source path are so weak,
that it is impossible to notice them in the plot.

\vfill

\noindent\parbox{12.7cm}{
\leftline {\bf OGLE 2005-BLG-017 (best fit)} 

\includegraphics[height=62mm,width=63mm]{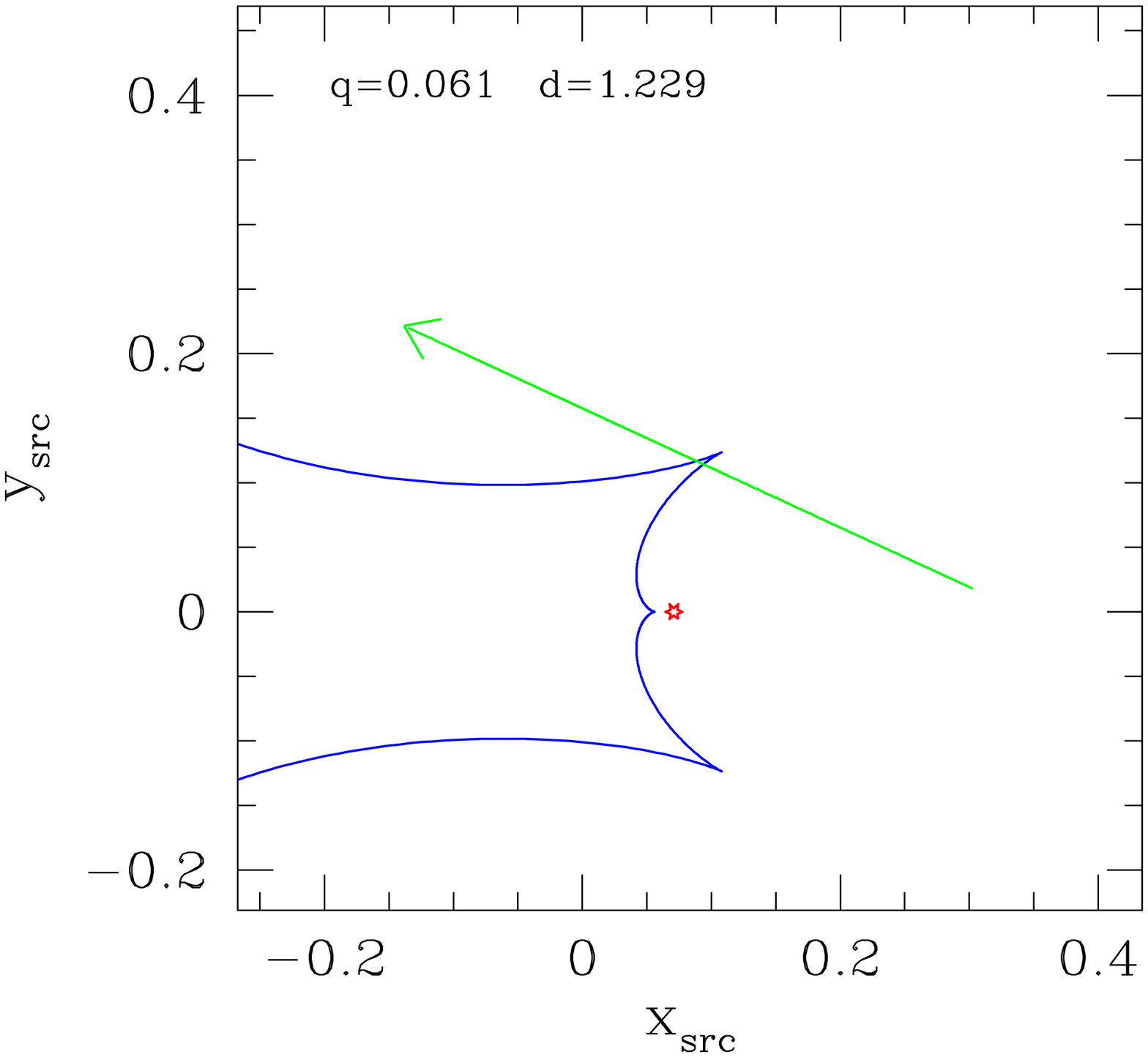}%
\includegraphics[height=62mm,width=63mm]{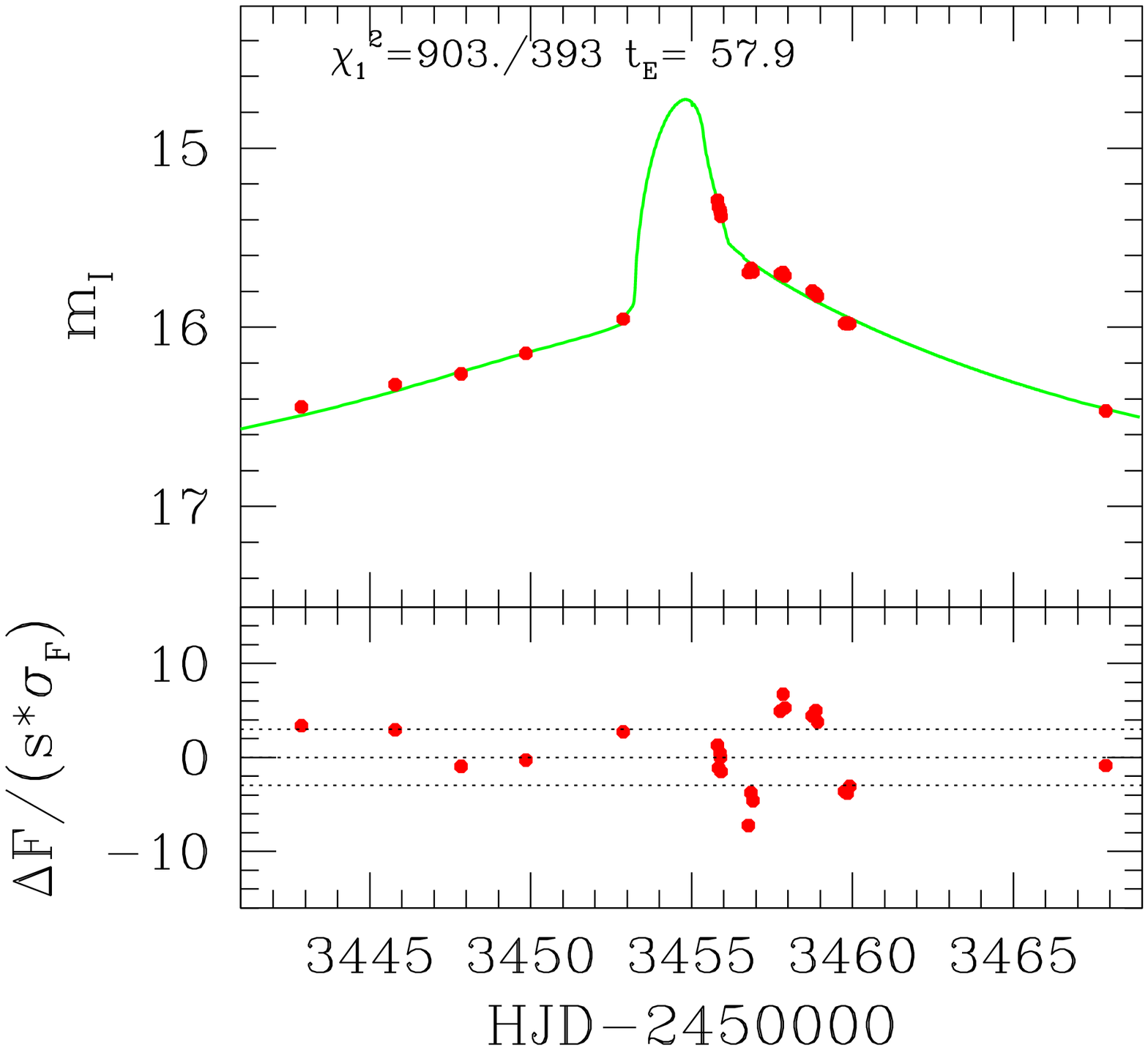}%

}

\noindent\parbox{12.7cm}{
\leftline {\bf OGLE 2005-BLG-017 (another model)} 

\includegraphics[height=62mm,width=63mm]{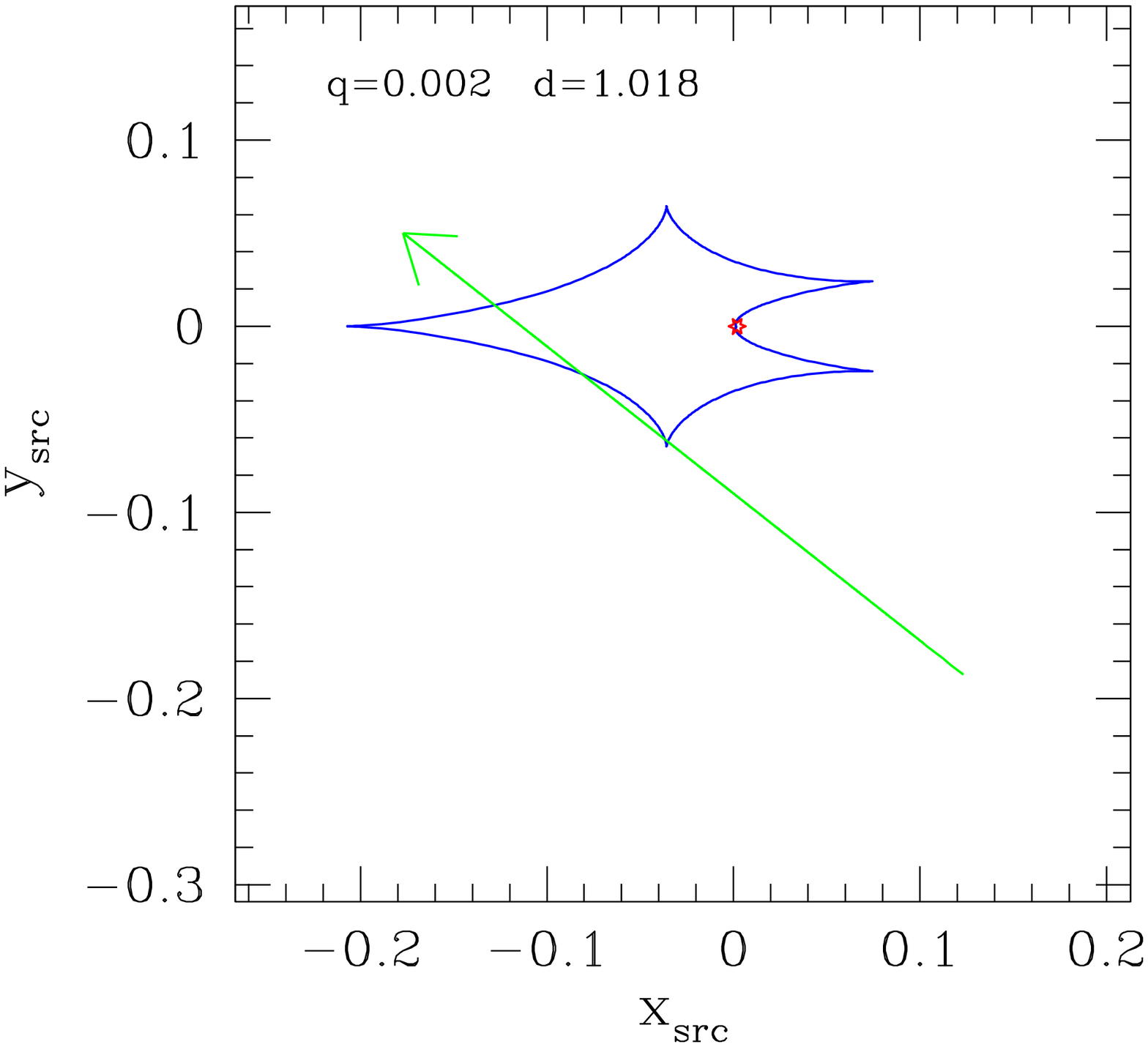}%
\includegraphics[height=62mm,width=63mm]{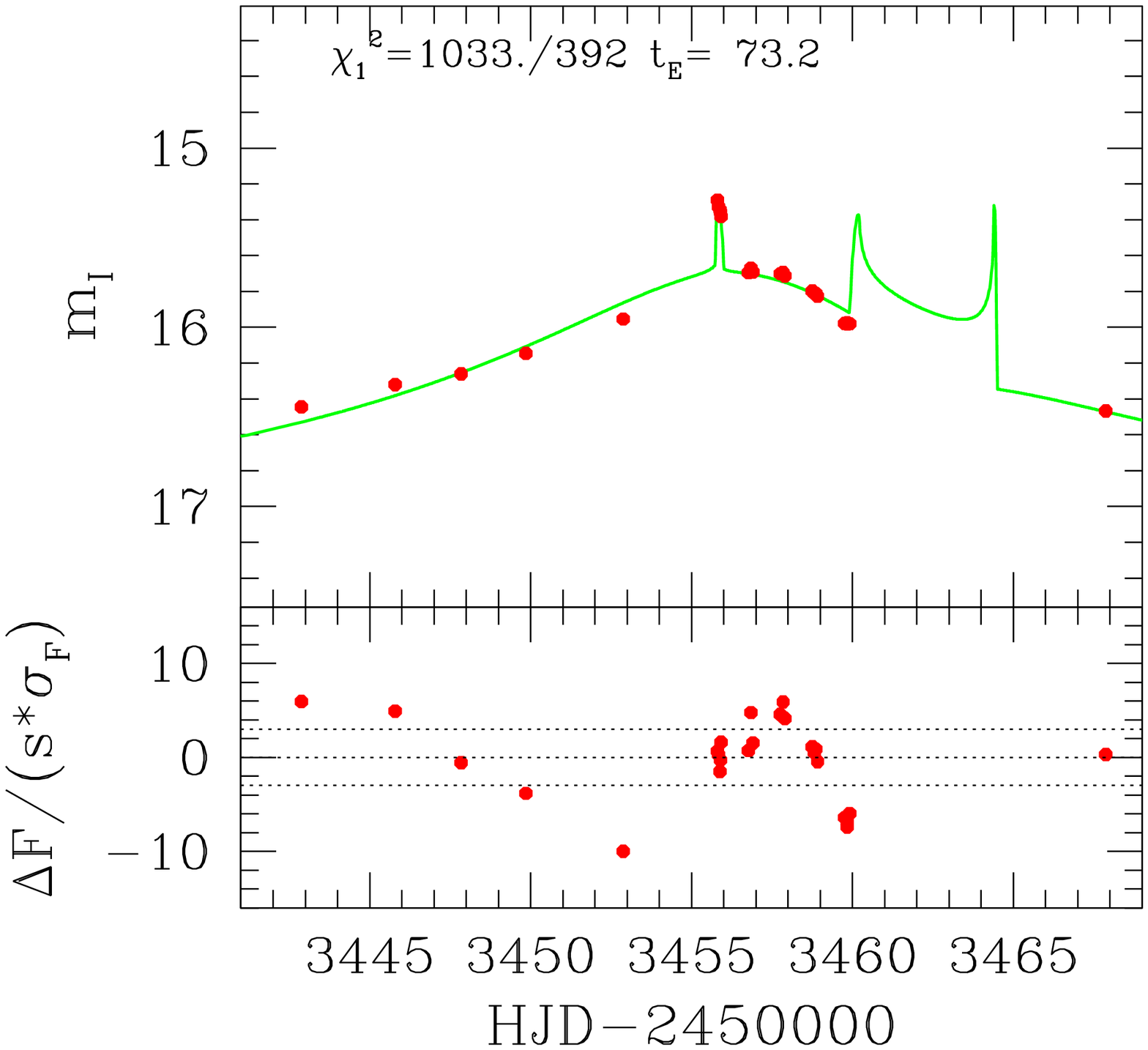}%

}

\noindent\parbox{12.75cm}{
\leftline {\bf OGLE 2005-BLG-018} 

\includegraphics[height=62mm]{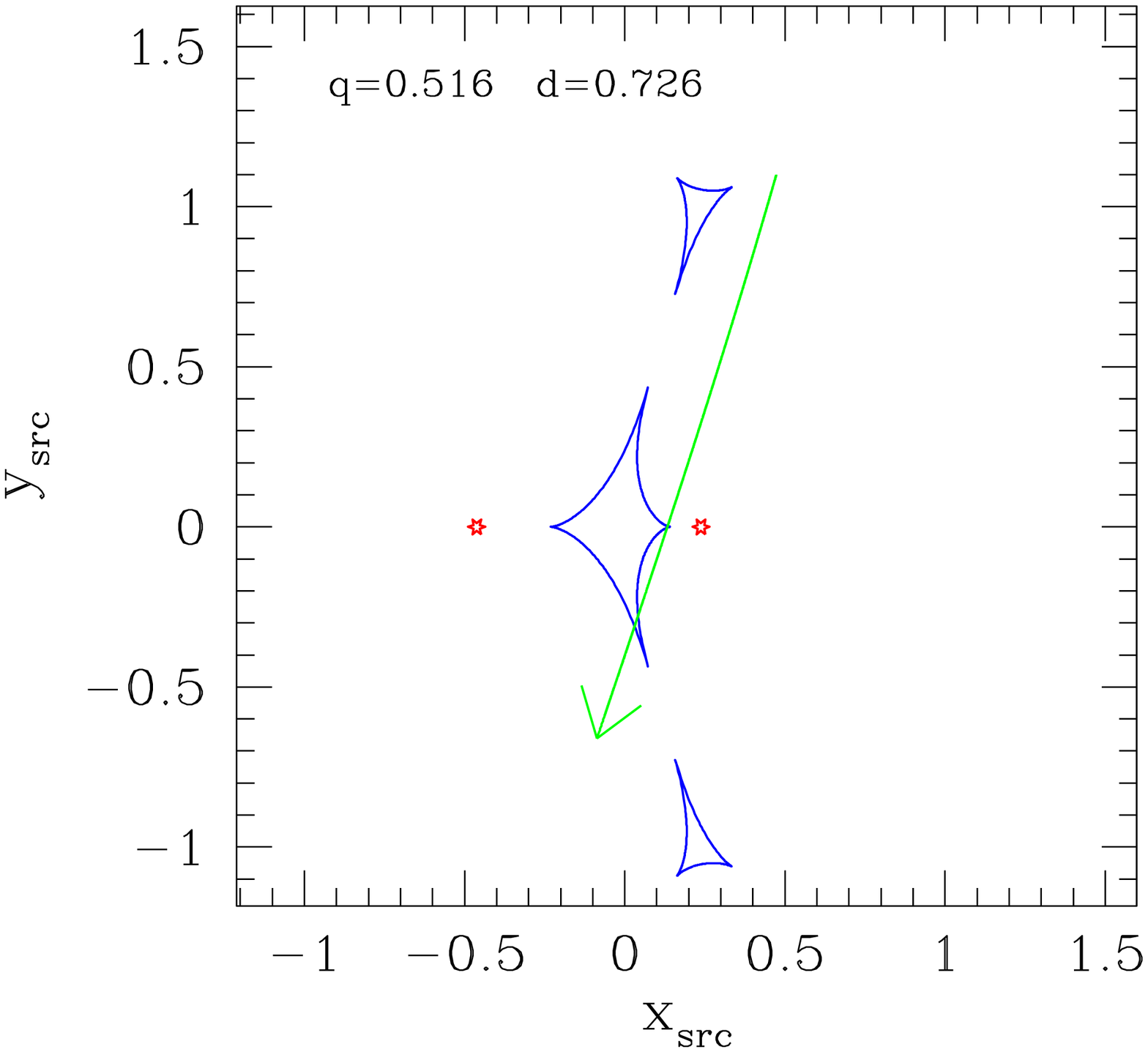}%
\includegraphics[height=62mm]{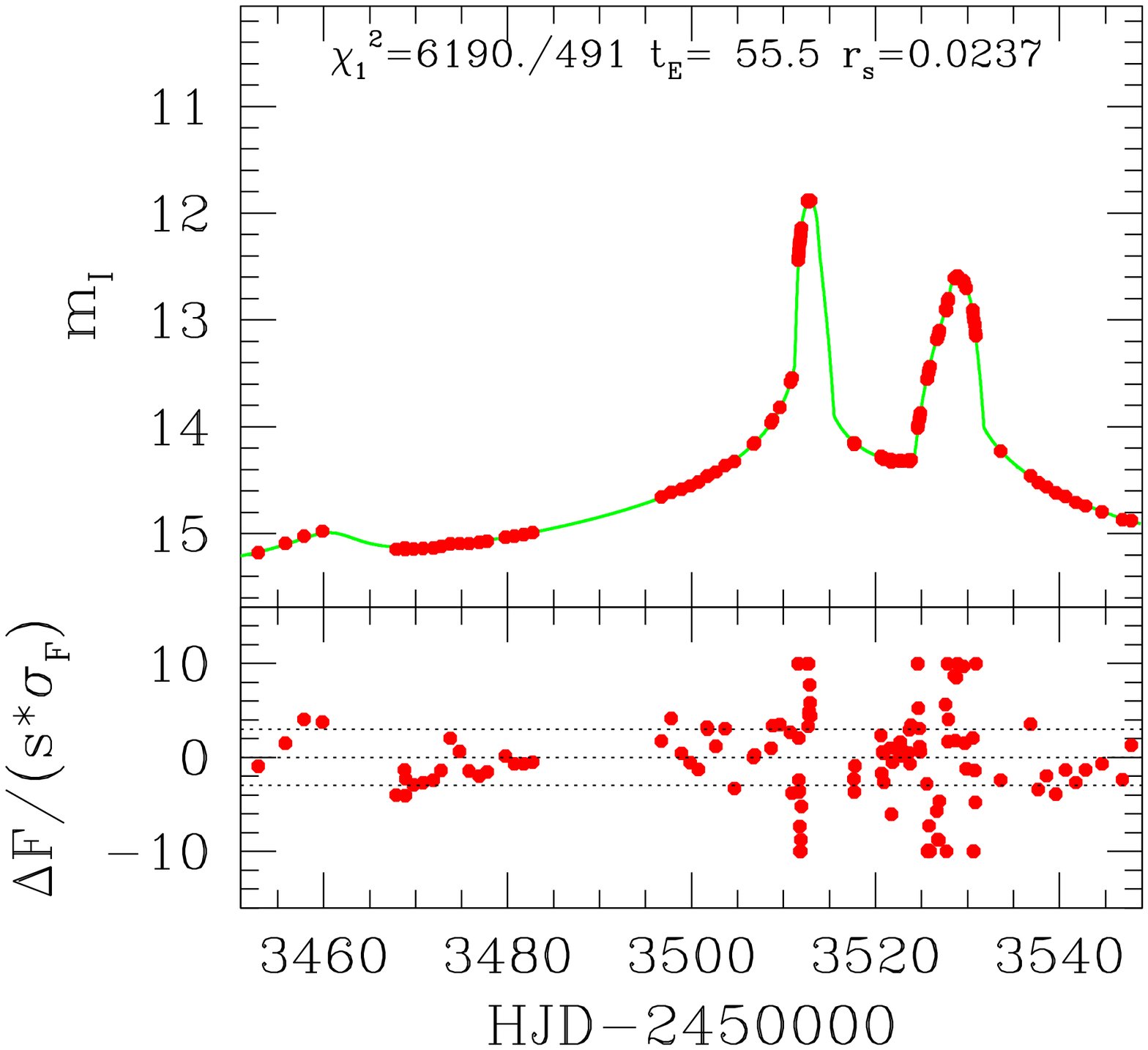}%

}

\noindent\parbox{12.75cm}{
\leftline {\bf OGLE 2005-BLG-062} 

\includegraphics[height=62mm,width=63mm]{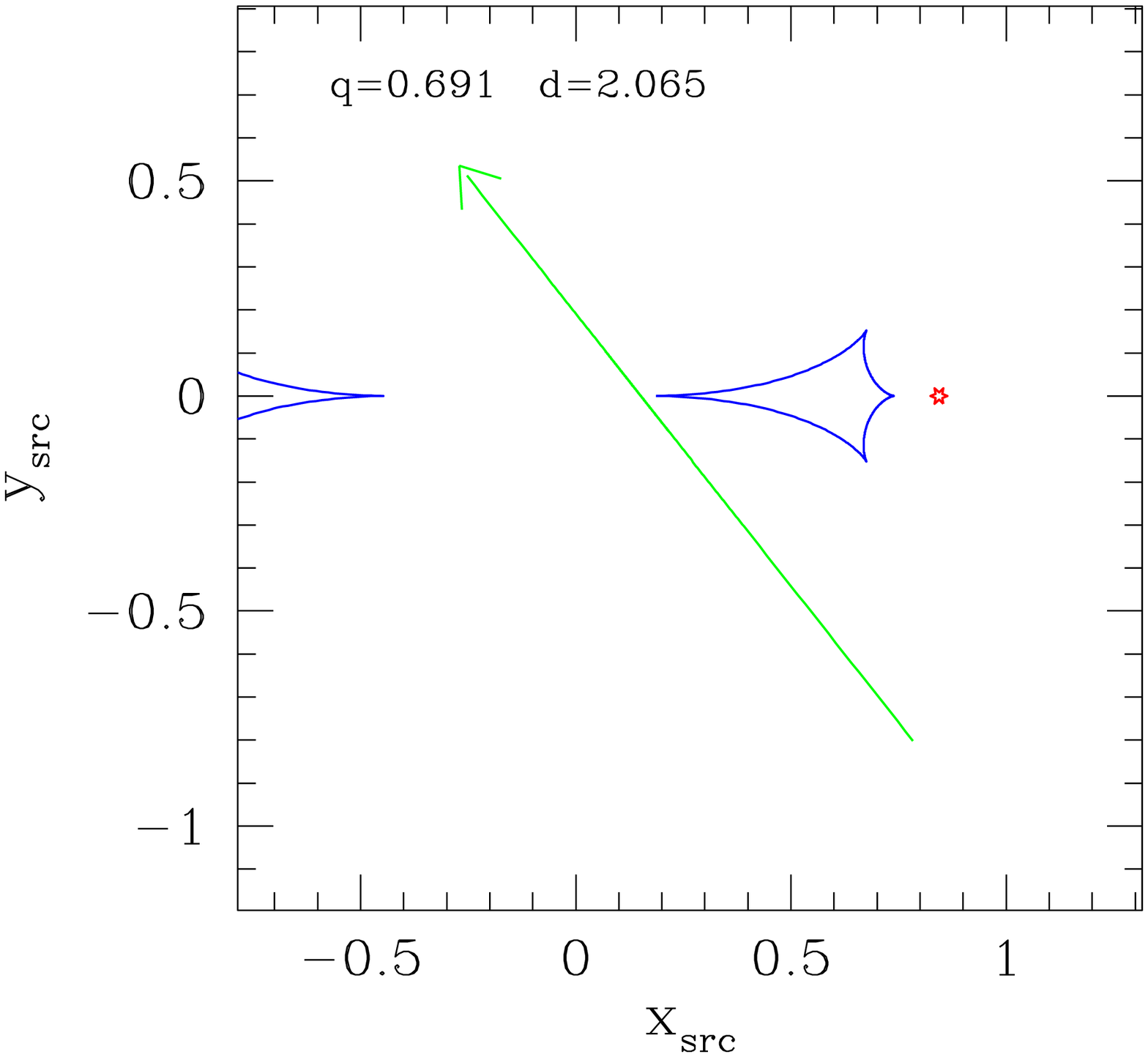}%
\includegraphics[height=62mm,width=63mm]{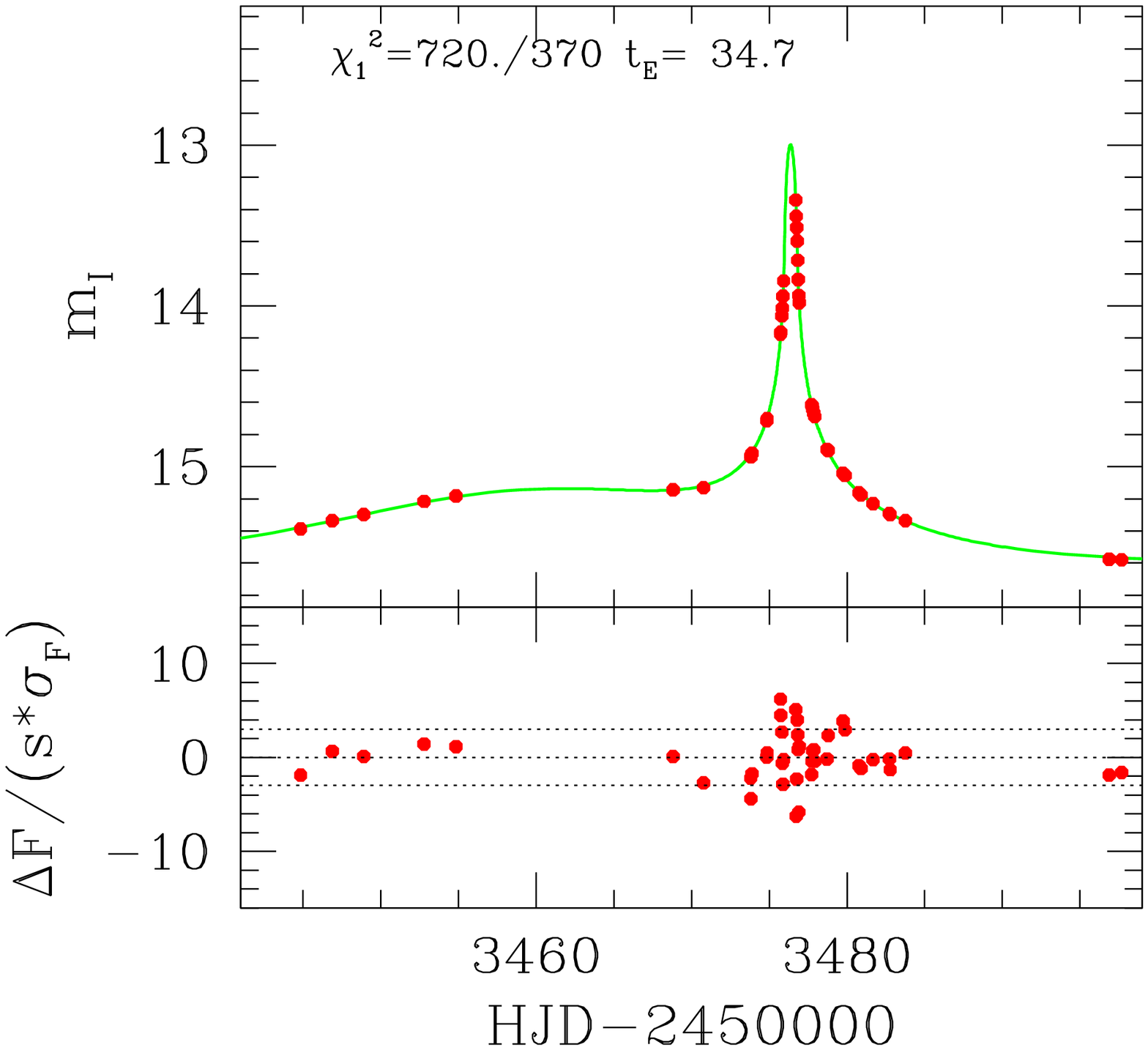}%

}

\noindent\parbox{12.75cm}{
\leftline {\bf OGLE 2005-BLG-128} 

 \includegraphics[height=62mm,width=63mm]{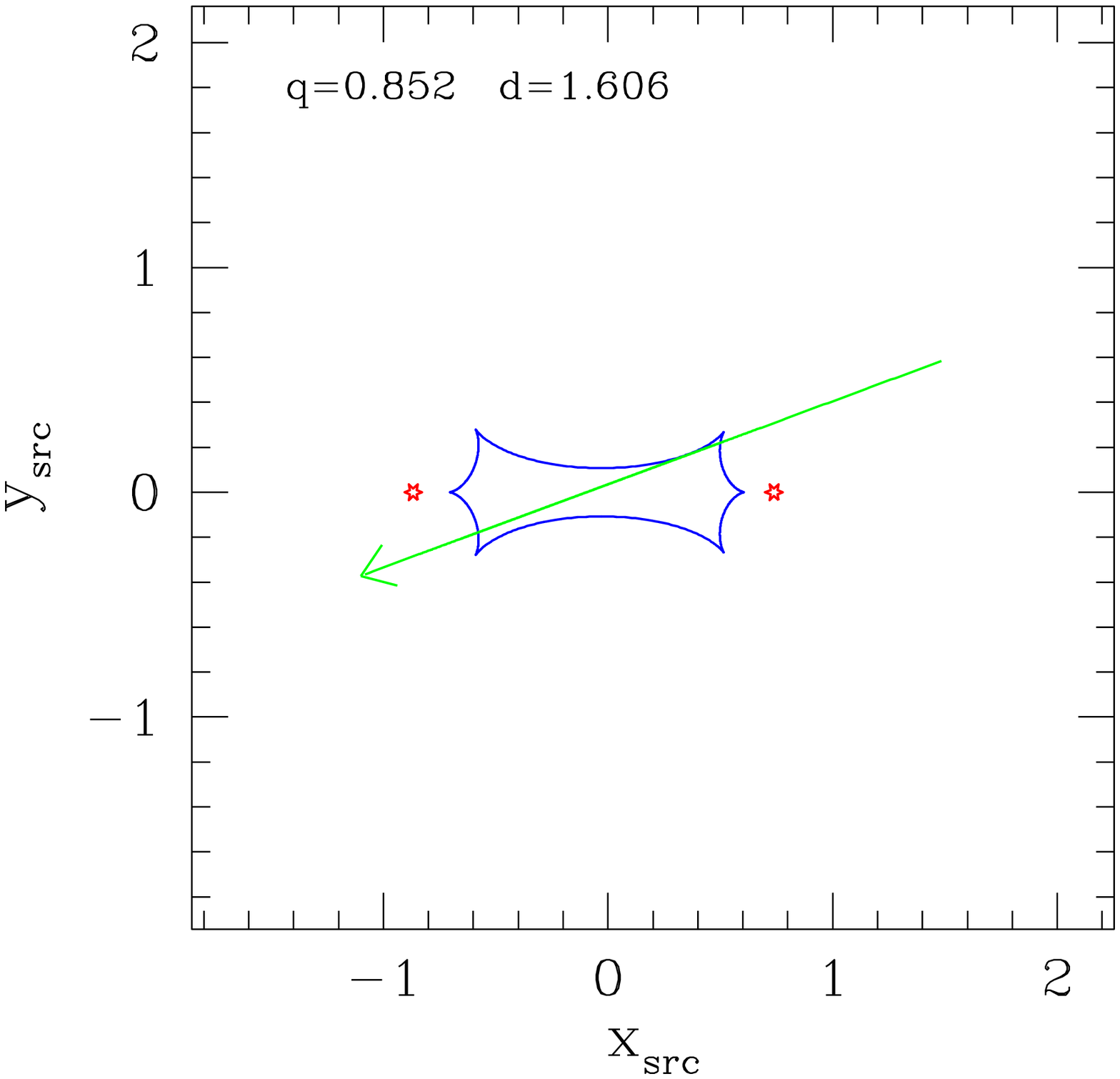}%
 \includegraphics[height=62mm,width=63mm]{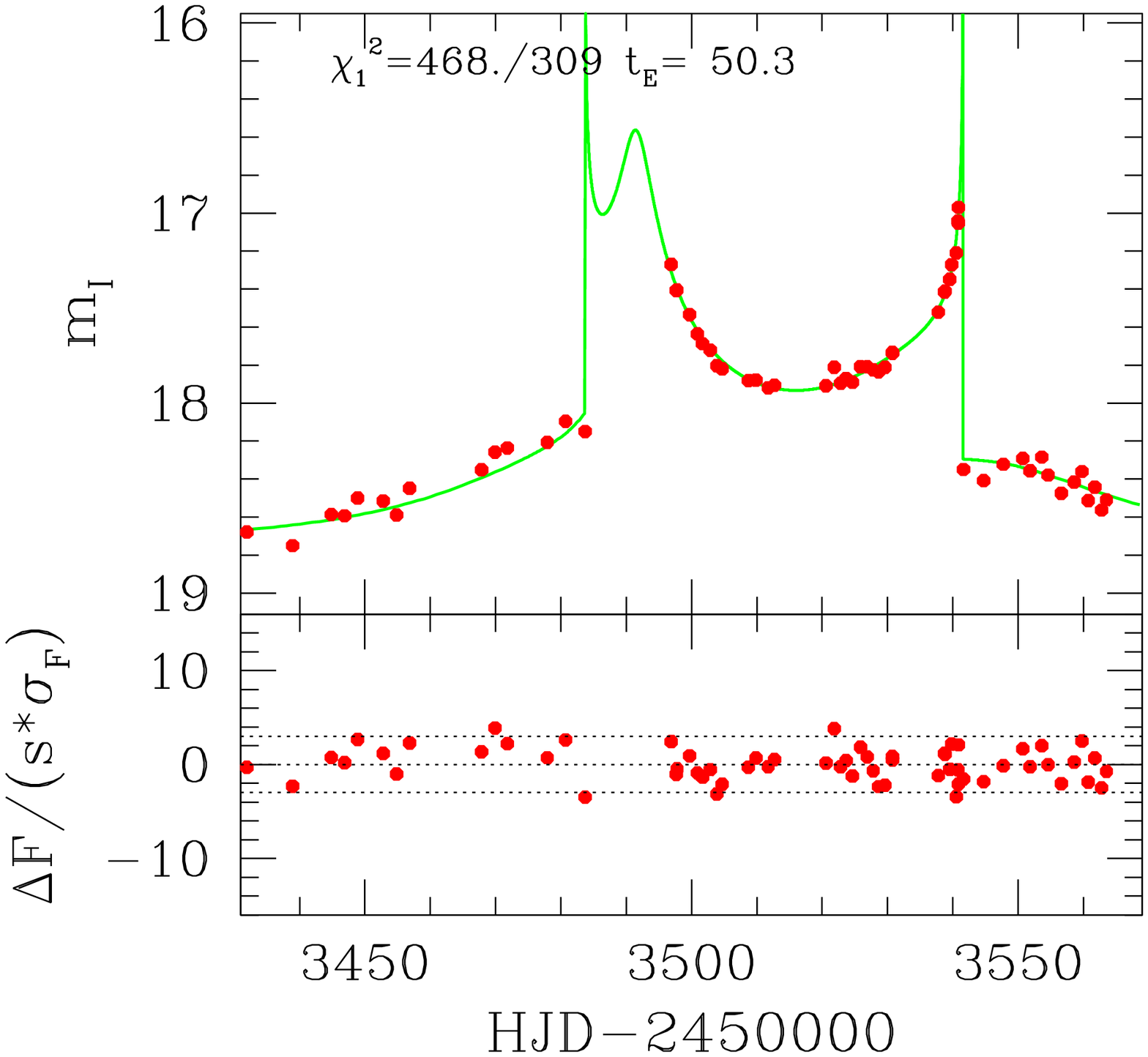}%

}

\noindent\parbox{12.75cm}{
\leftline {\bf OGLE 2005-BLG-153} 

 \includegraphics[height=62mm,width=63mm]{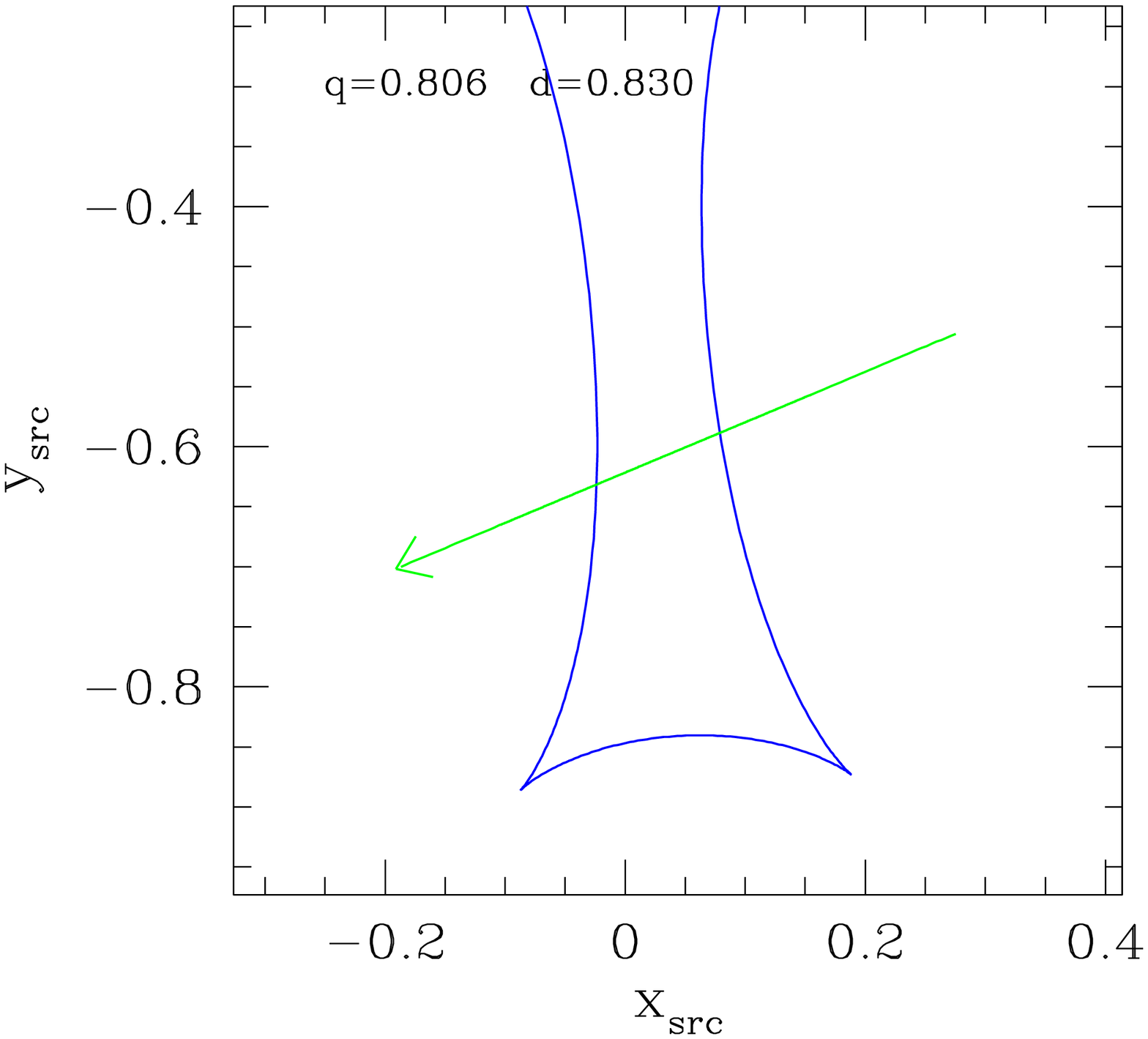}%
 \includegraphics[height=62mm,width=63mm]{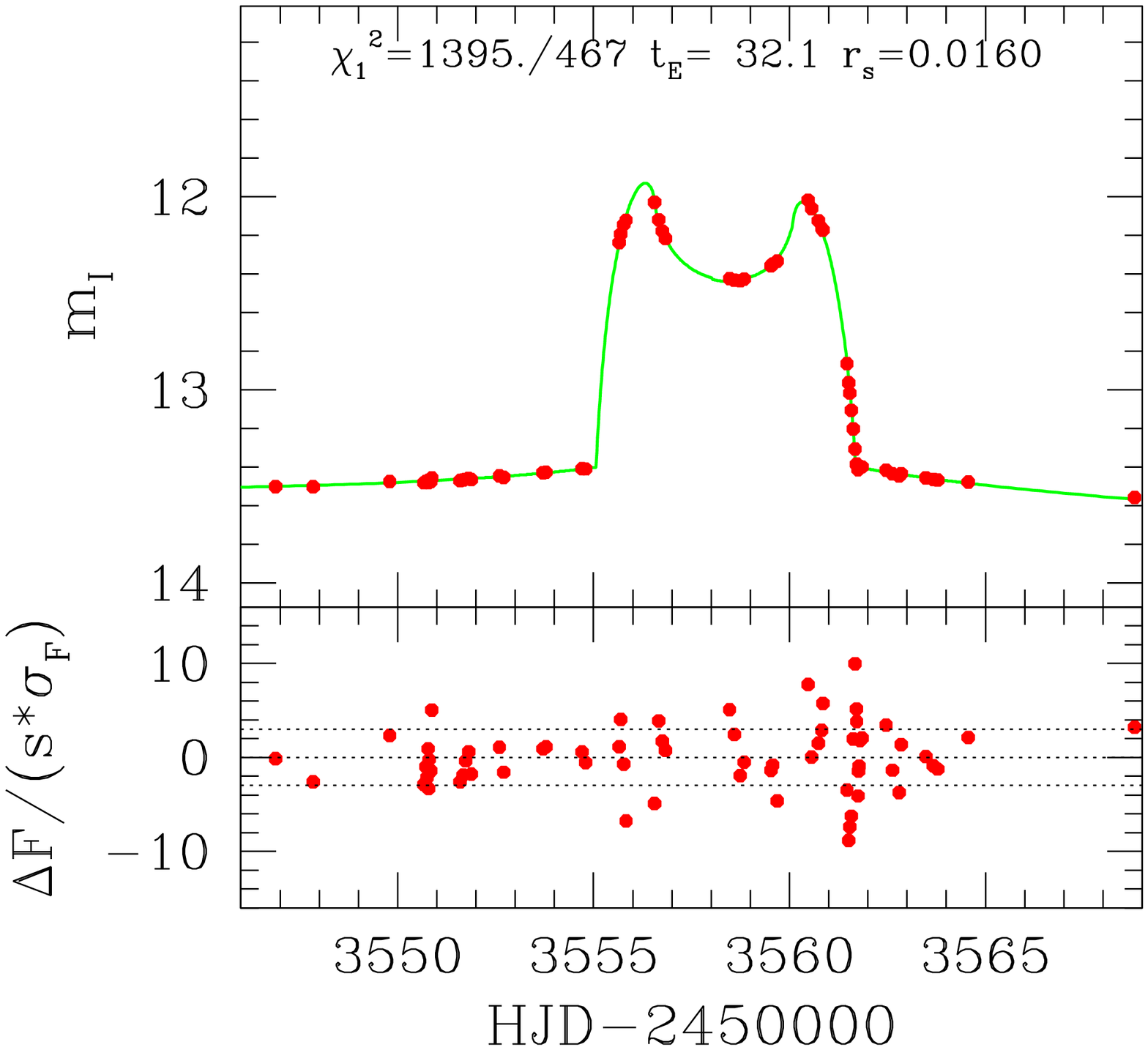}%

}

\noindent\parbox{12.75cm}{
\leftline {\bf OGLE 2005-BLG-189} 

\includegraphics[height=62mm,width=63mm]{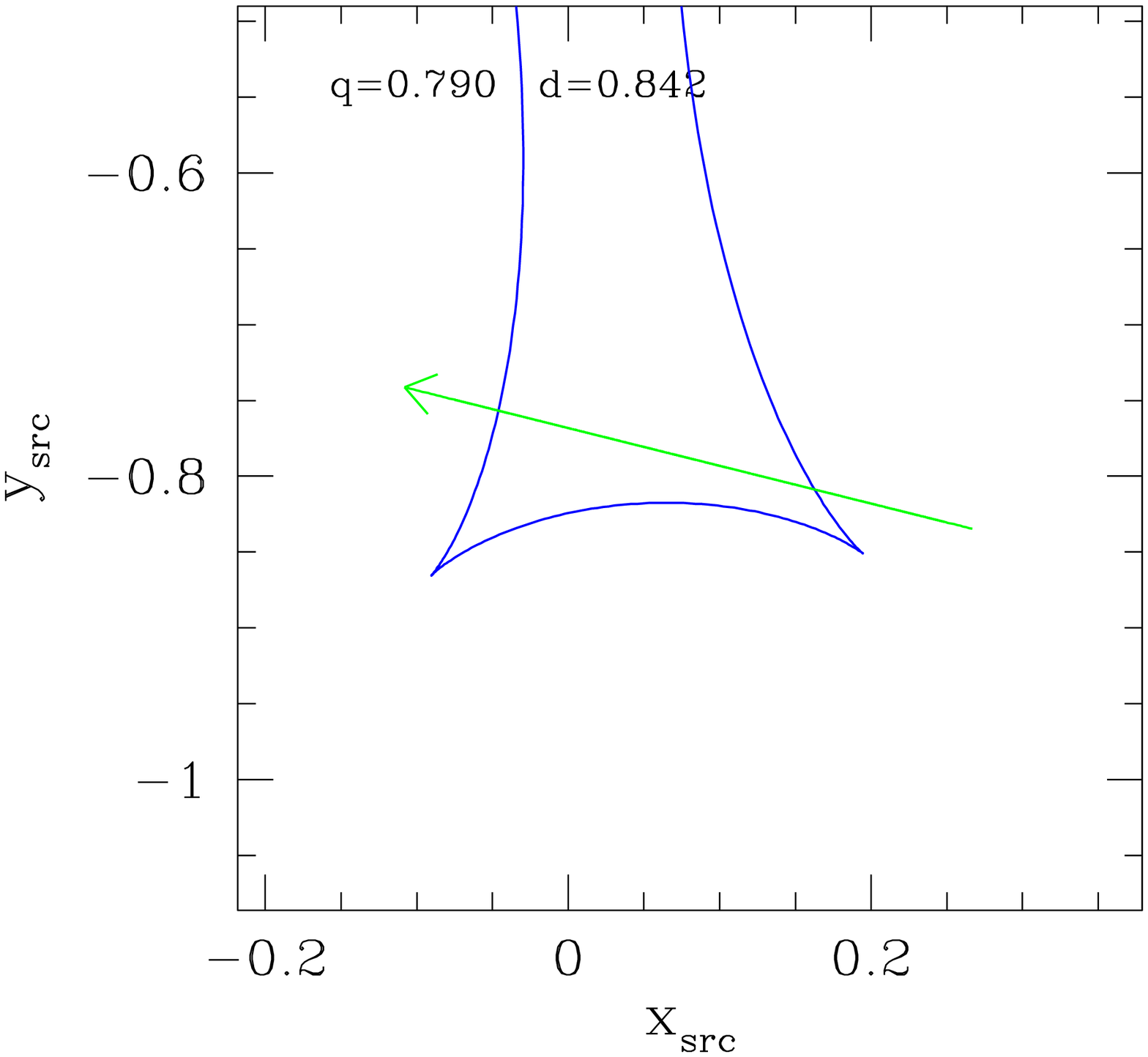}%
\includegraphics[height=62mm,width=63mm]{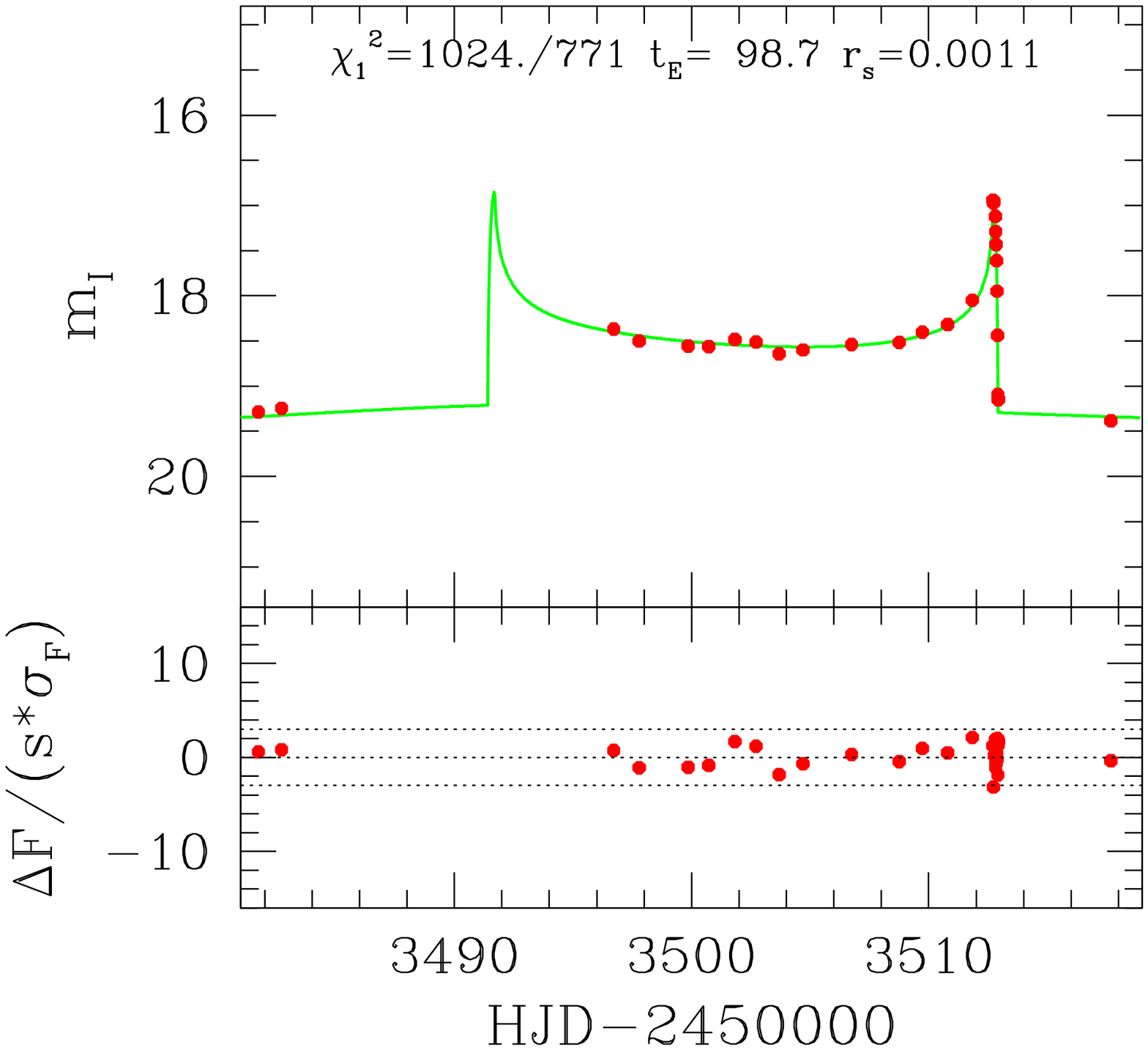}%

}

\noindent\parbox{12.75cm}{
\leftline {\bf OGLE 2005-BLG-226} 

\includegraphics[height=62mm,width=63mm]{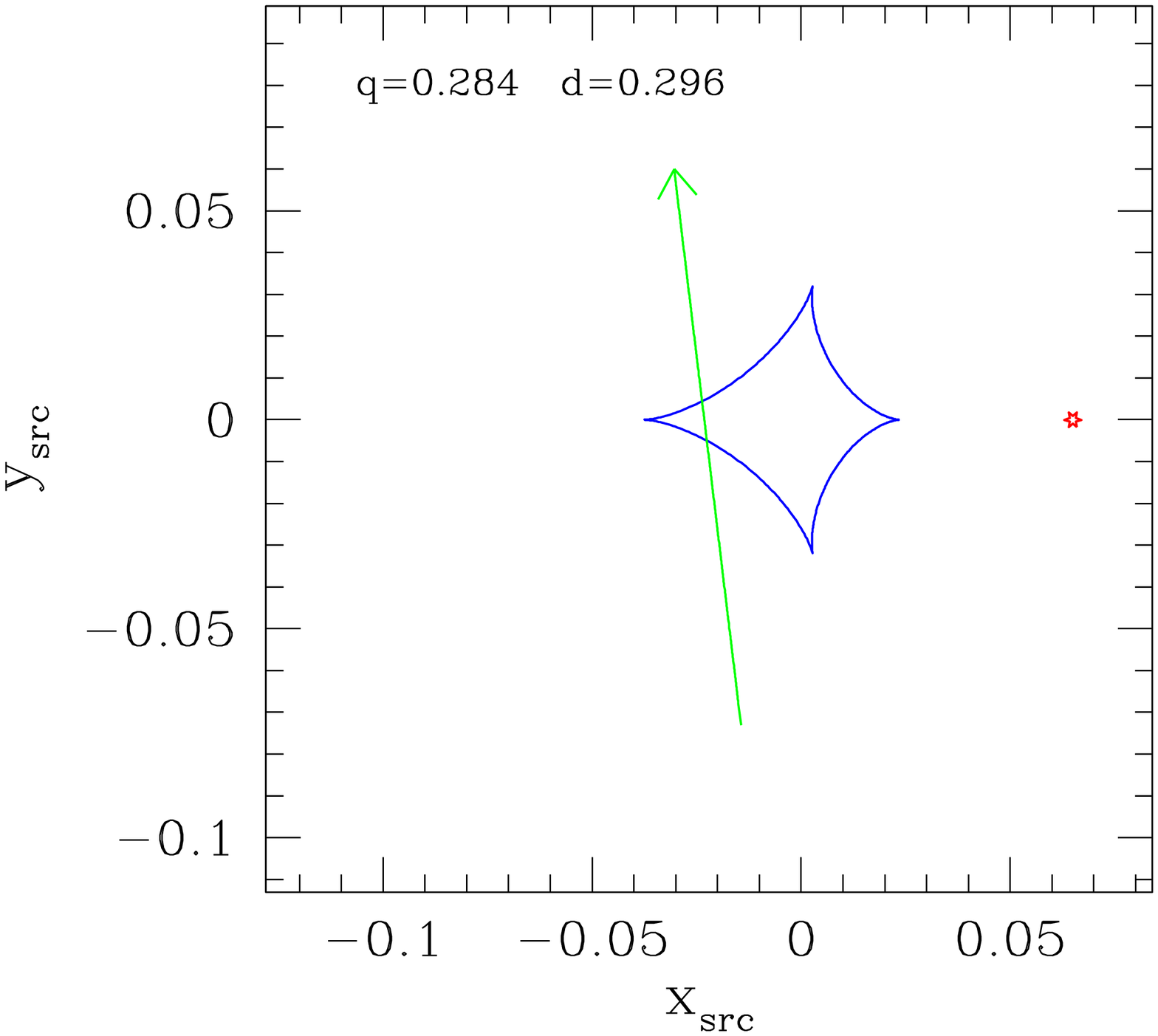}%
\includegraphics[height=62mm,width=63mm]{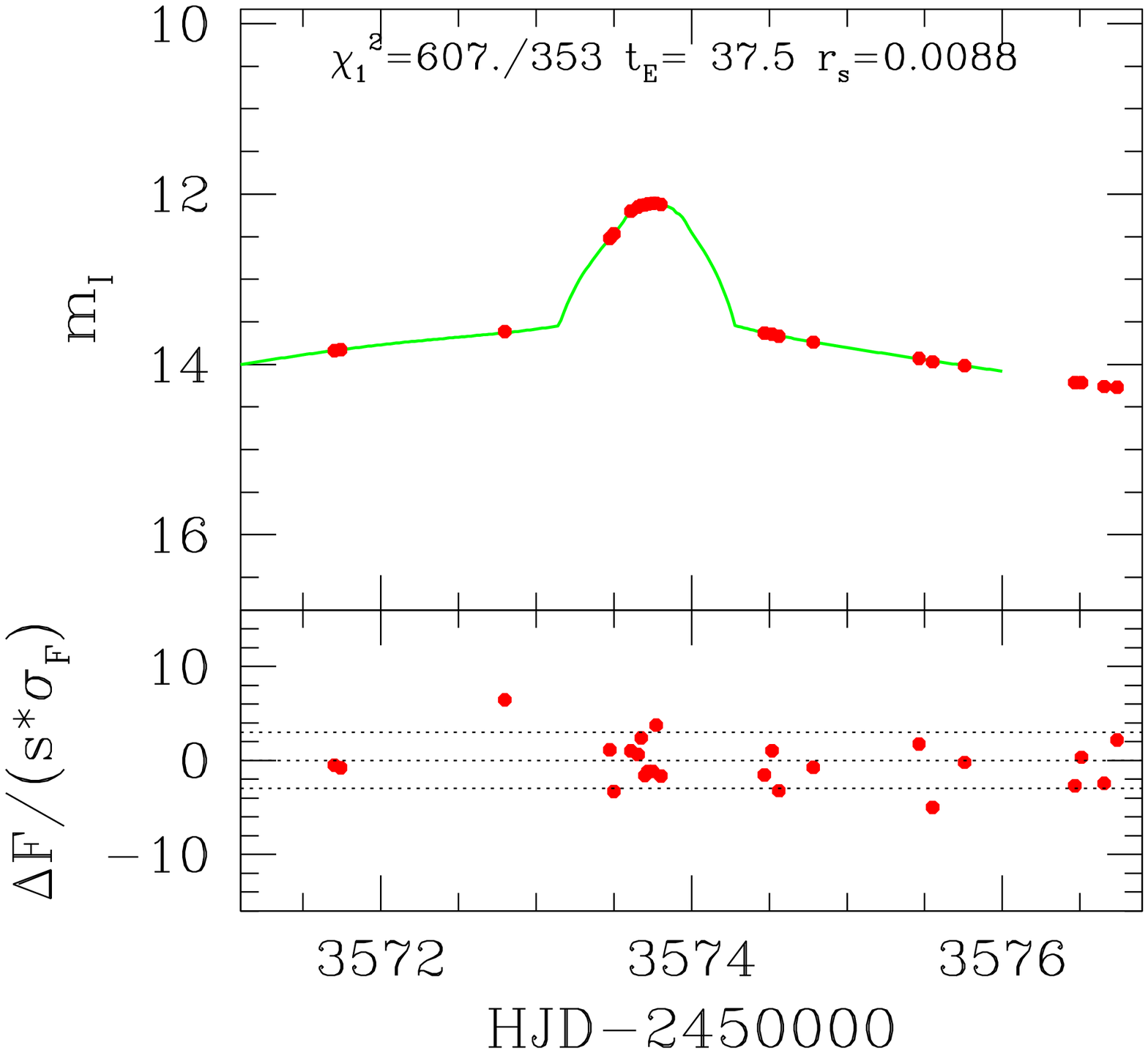}%

}

\noindent\parbox{12.75cm}{
\leftline {\bf OGLE 2005-BLG-327 (close)} 

 \includegraphics[height=62mm,width=63mm]{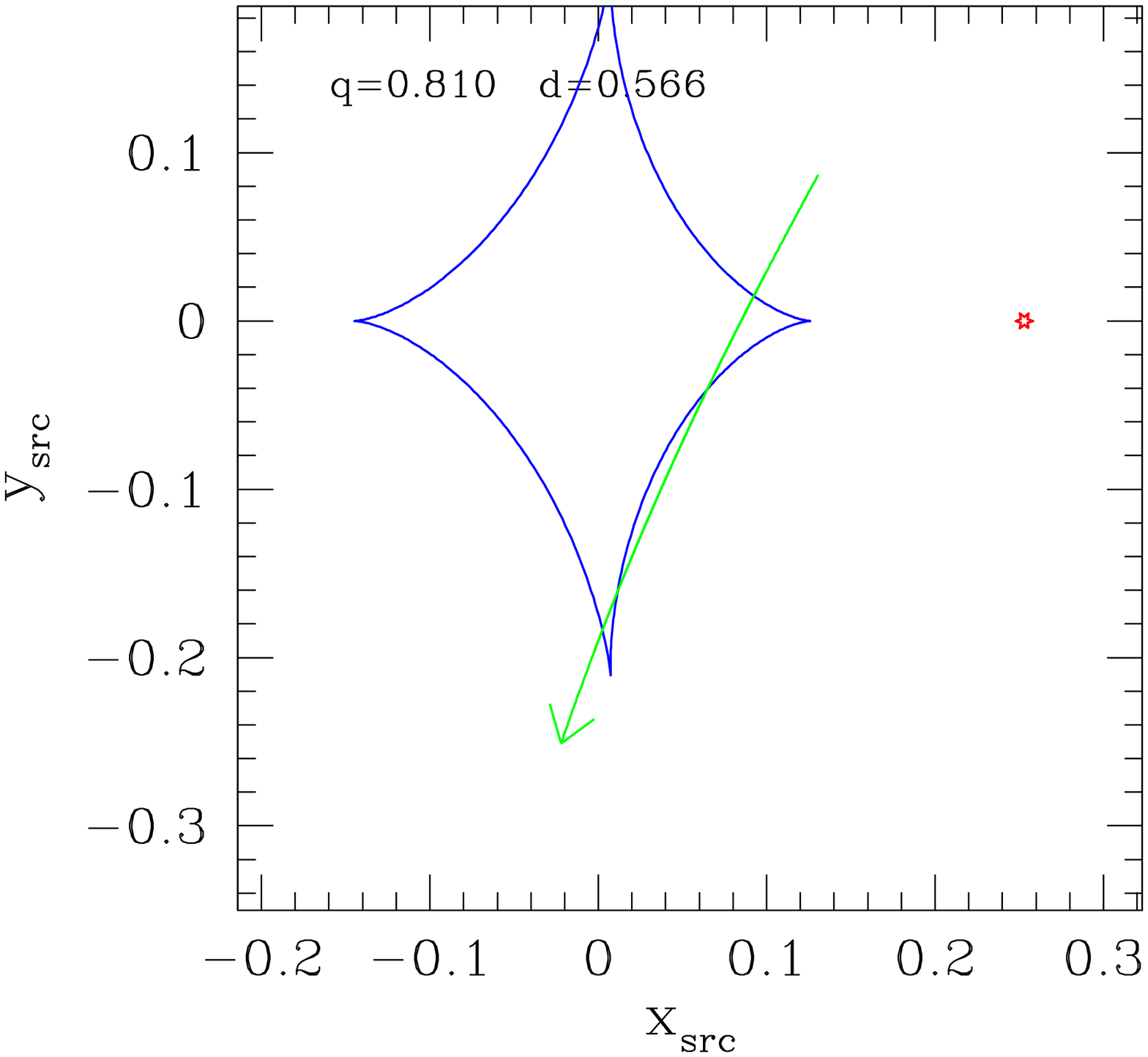}%
 \includegraphics[height=62mm,width=63mm]{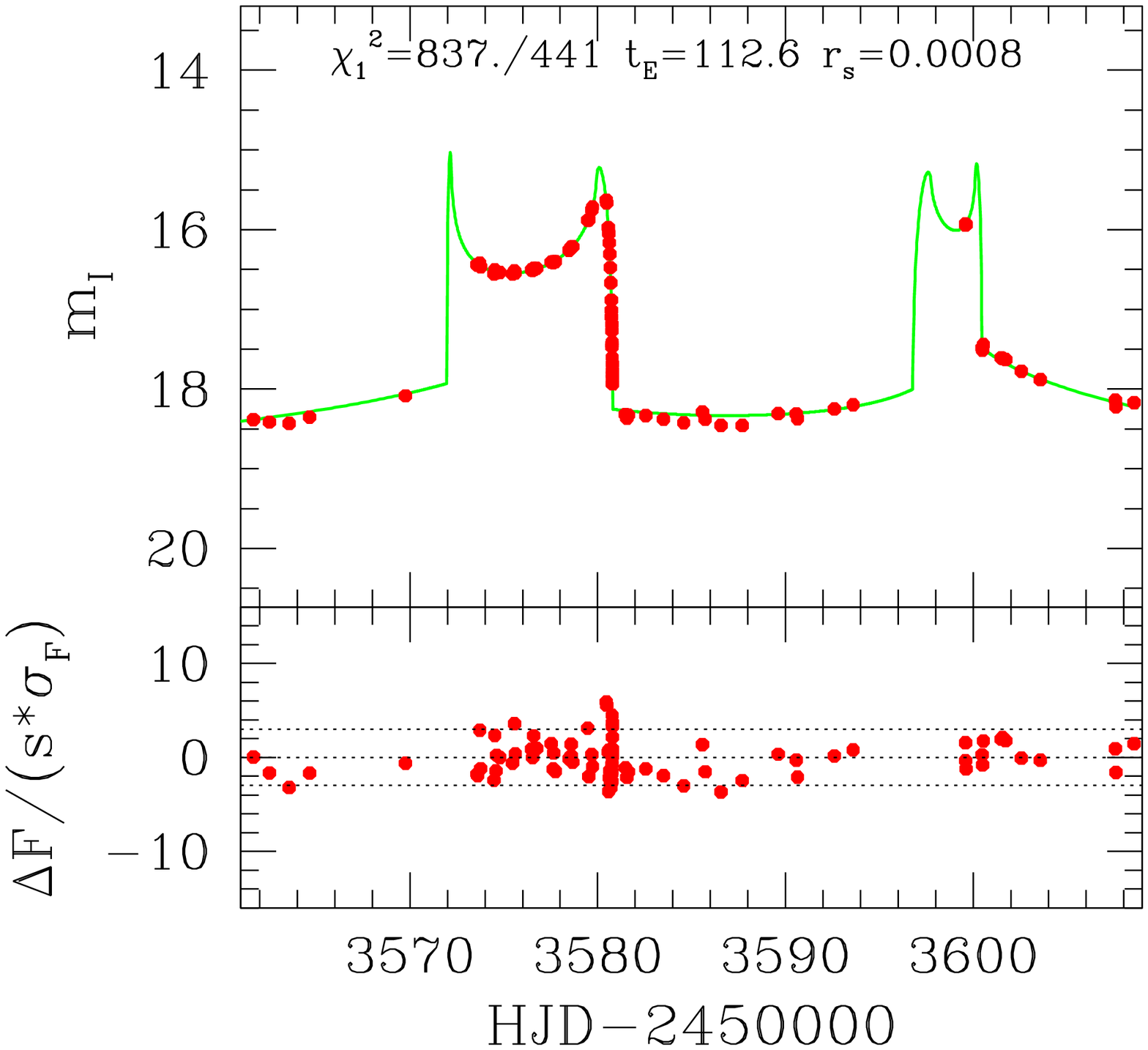}%

}

\noindent\parbox{12.75cm}{
\leftline {\bf OGLE 2005-BLG-327 (wide)} 

 \includegraphics[height=62mm,width=63mm]{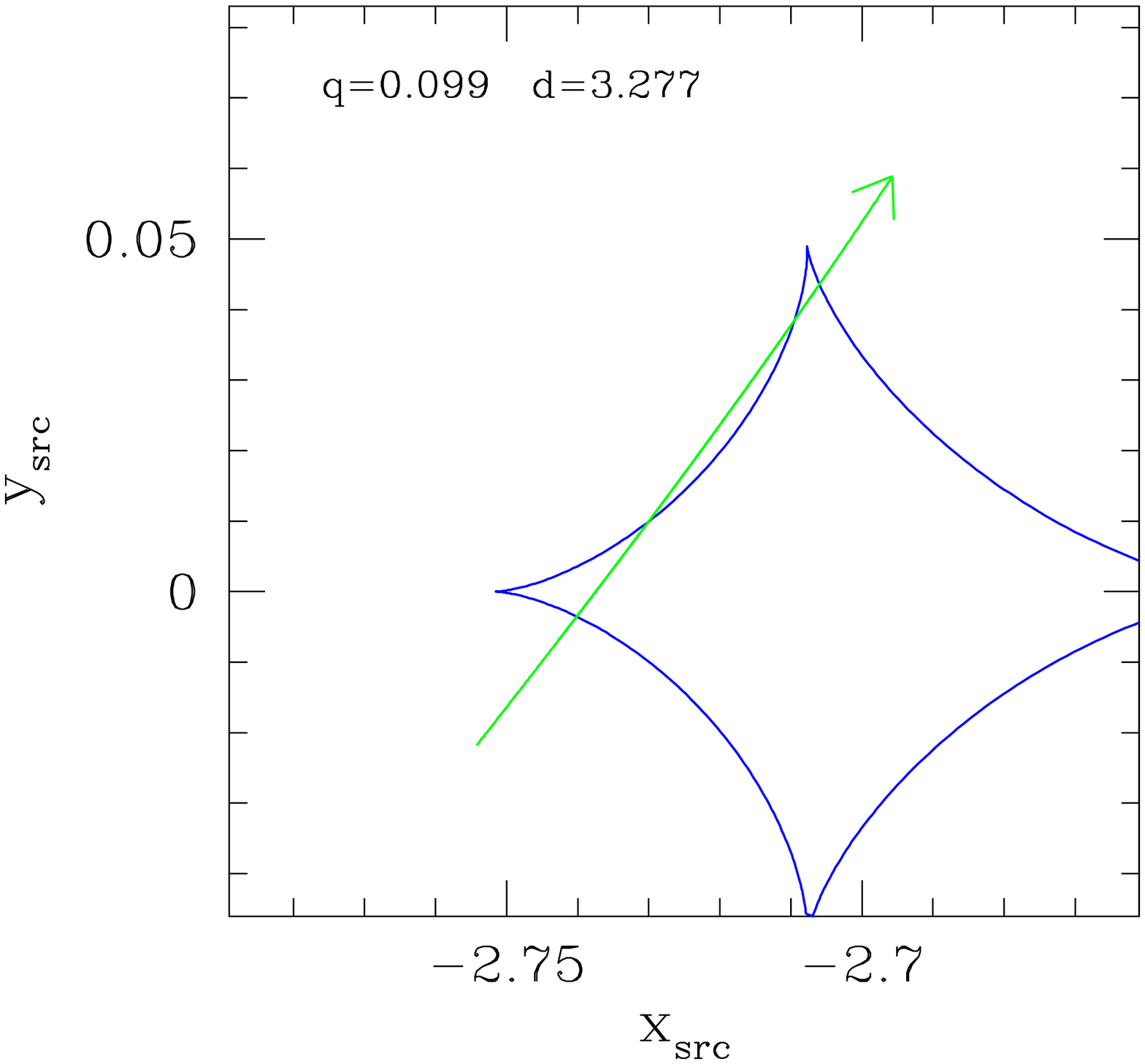}%
 \includegraphics[height=62mm,width=63mm]{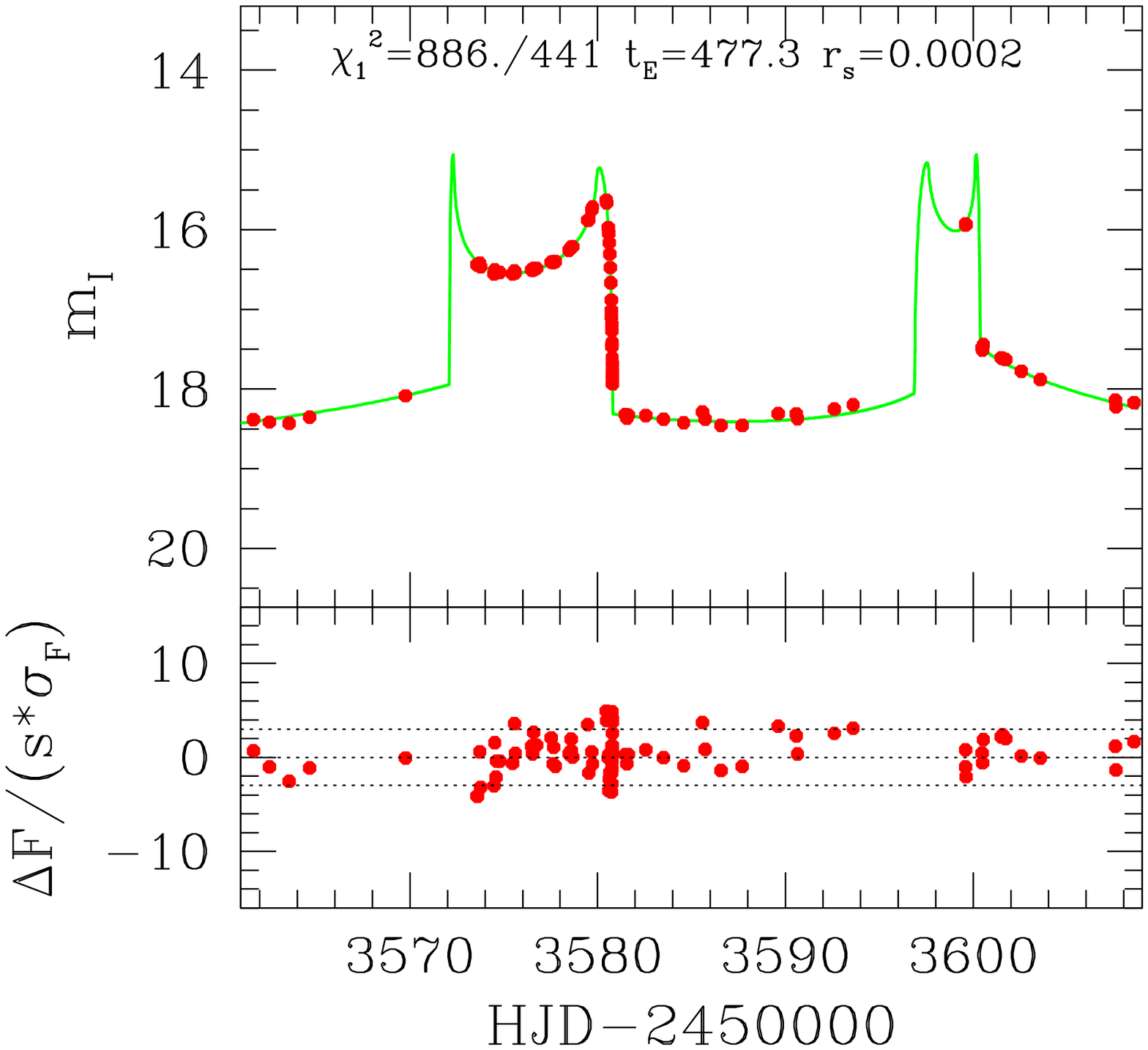}%

}

\noindent\parbox{12.75cm}{
\leftline {\bf OGLE 2005-BLG-331} 

 \includegraphics[height=62mm,width=63mm]{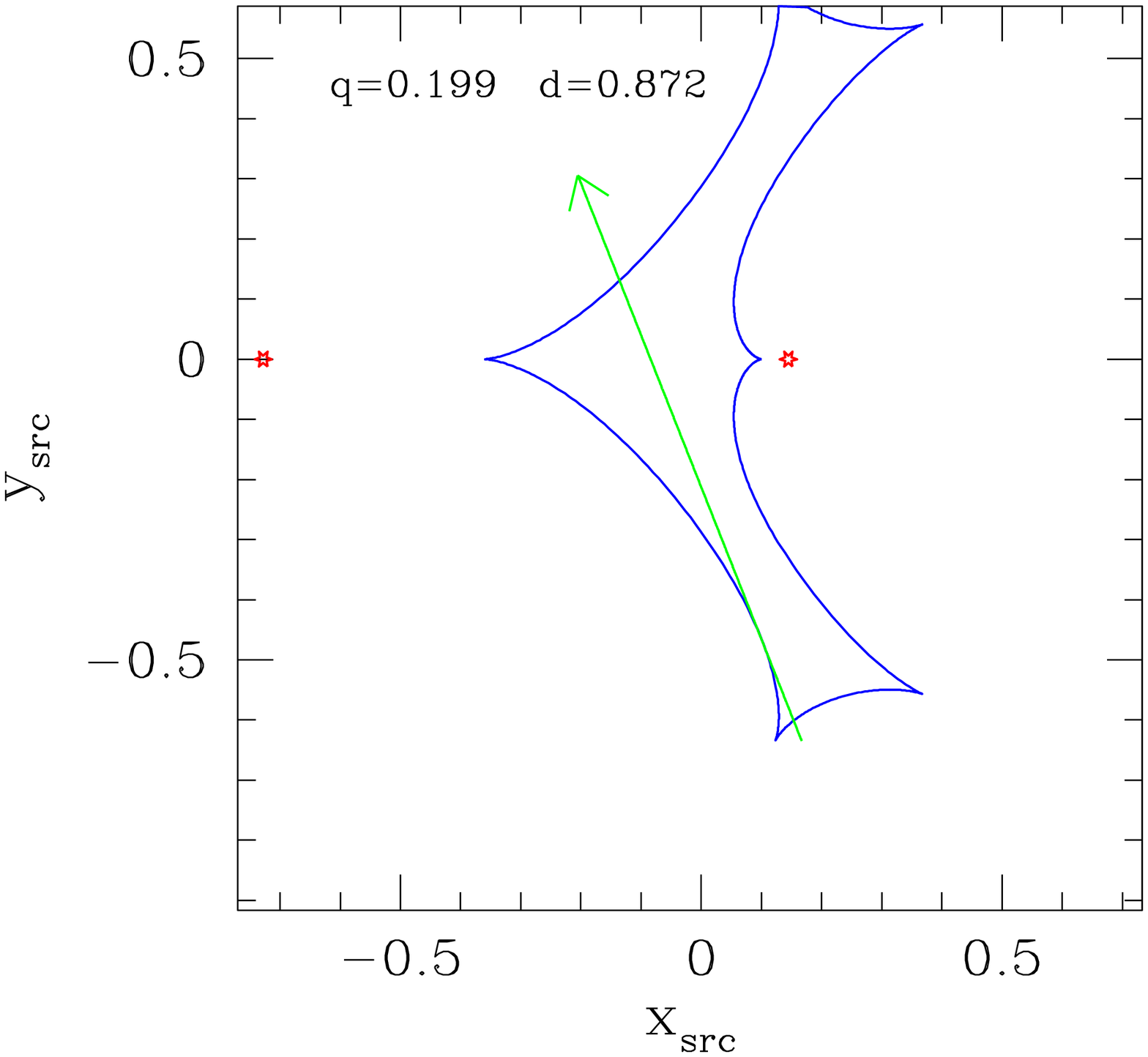}%
 \includegraphics[height=62mm,width=63mm]{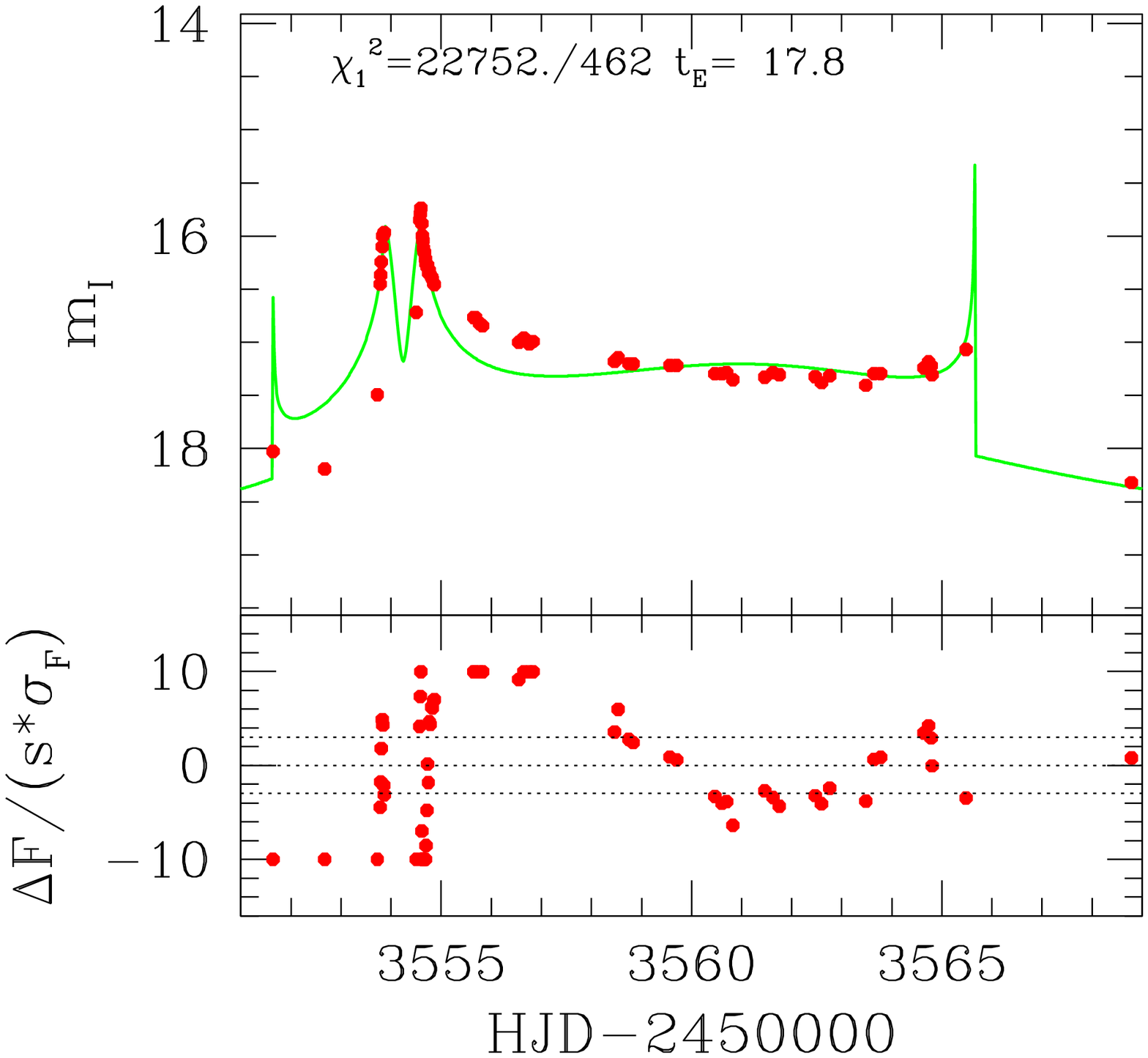}%

}

\noindent\parbox{12.75cm}{
\leftline {\bf OGLE 2005-BLG-463} 

 \includegraphics[height=62mm,width=63mm]{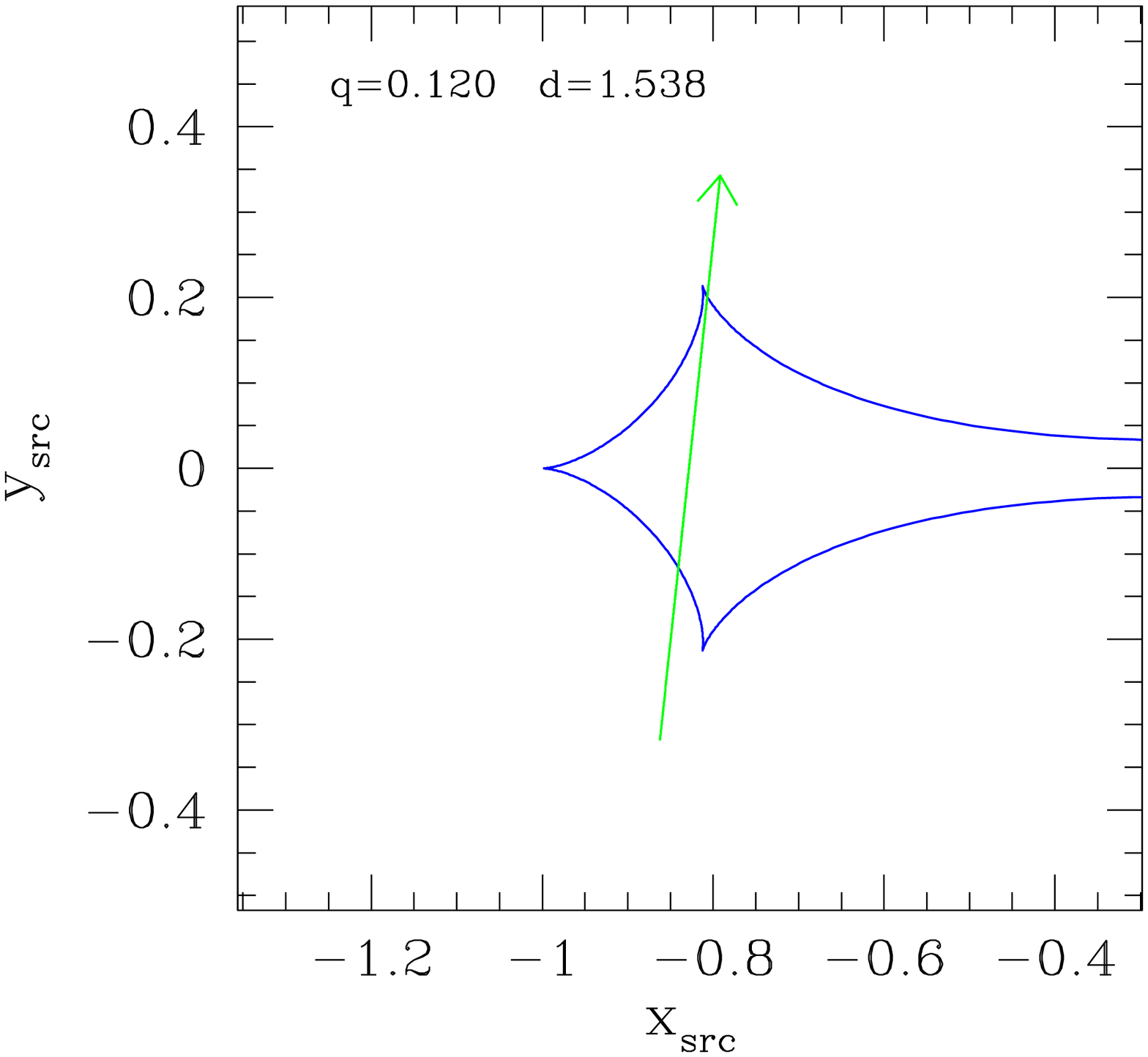}%
 \includegraphics[height=62mm,width=63mm]{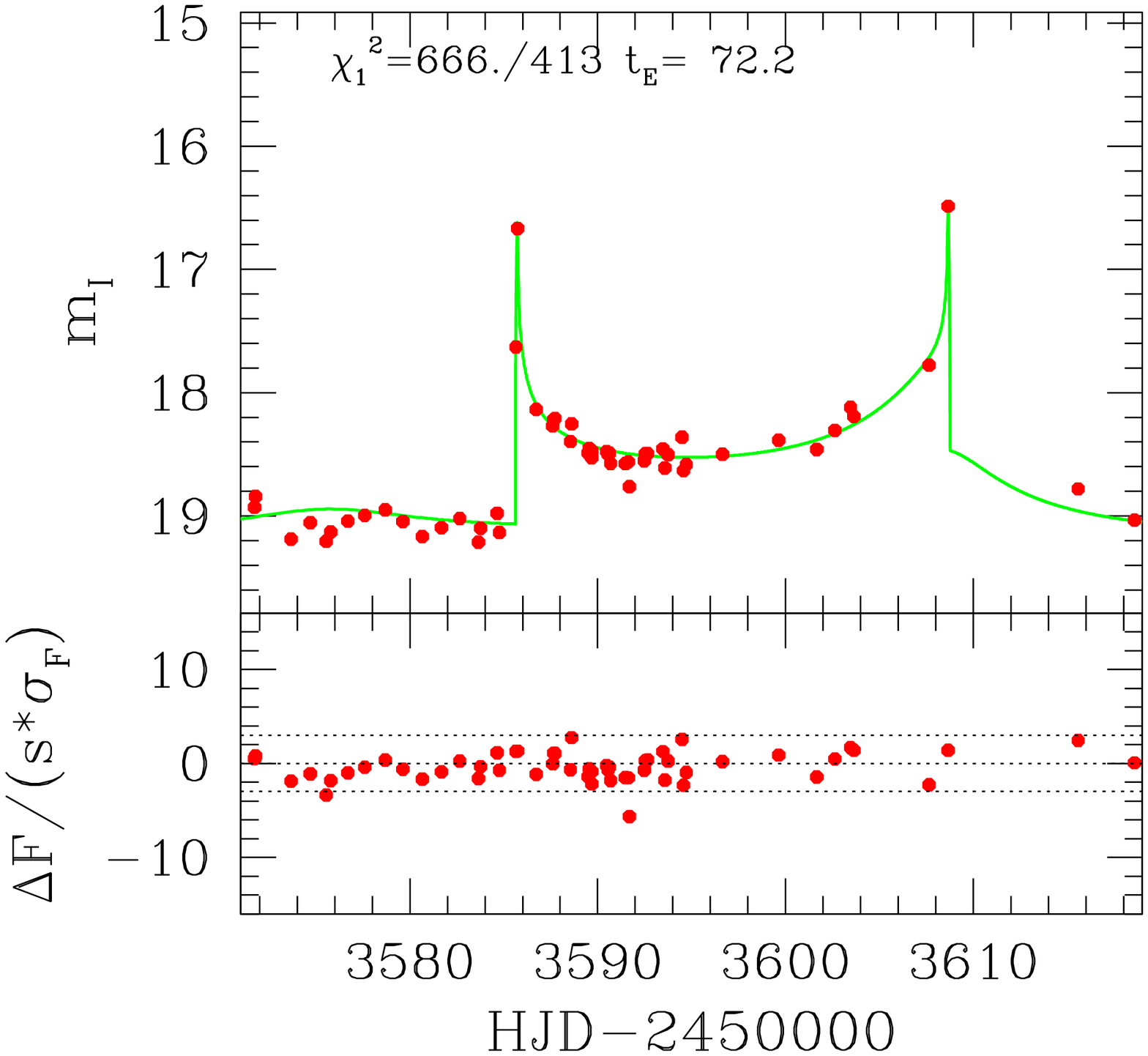}%

}

\noindent\parbox{12.75cm}{
\leftline {\bf OGLE 2005-BLG-468} 

 \includegraphics[height=62mm,width=63mm]{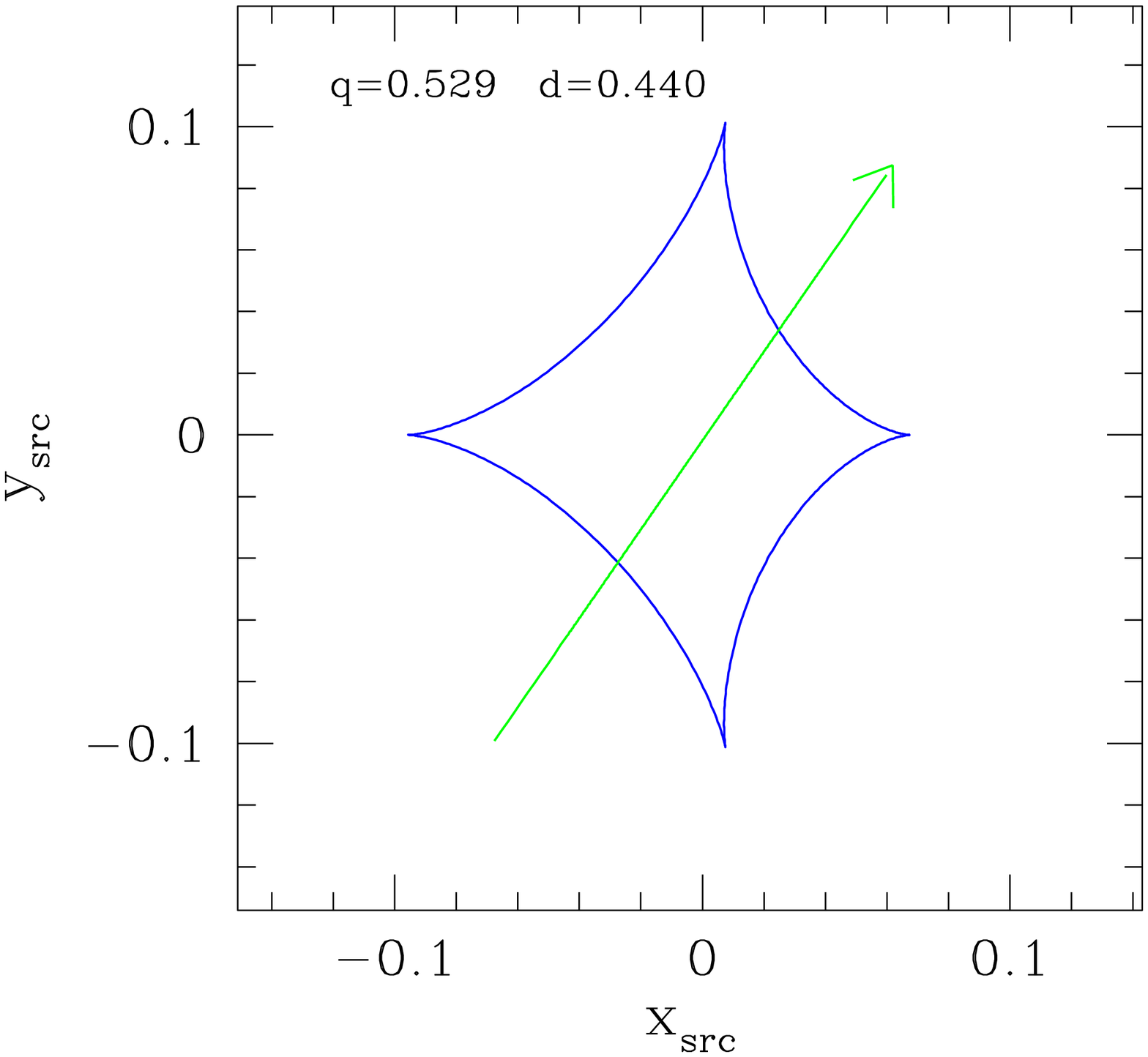}%
 \includegraphics[height=62mm,width=63mm]{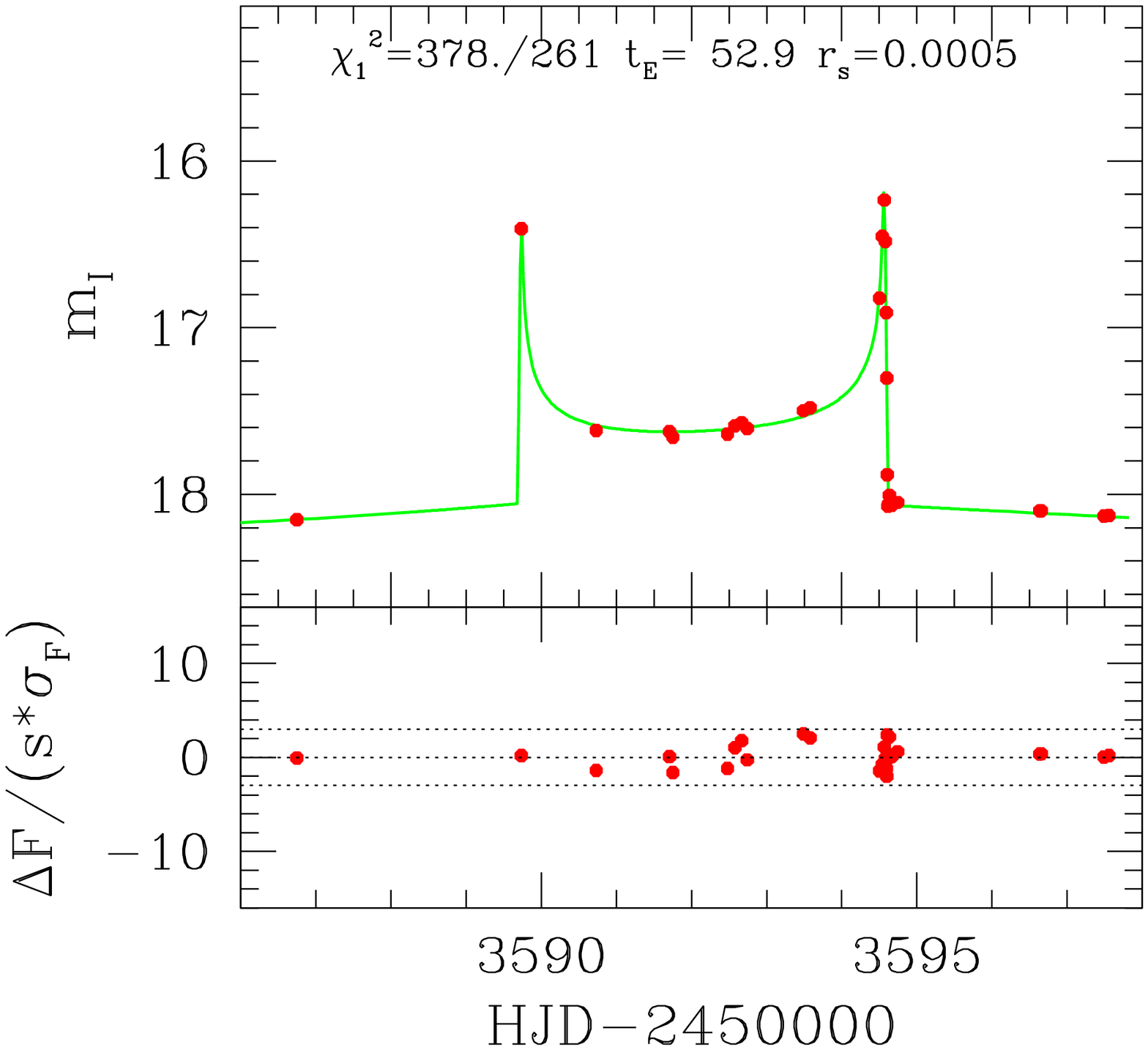}%

}

\noindent\parbox{12.75cm}{
\leftline {\bf OGLE 2005-BLG-477} 

 \includegraphics[height=62mm,width=63mm]{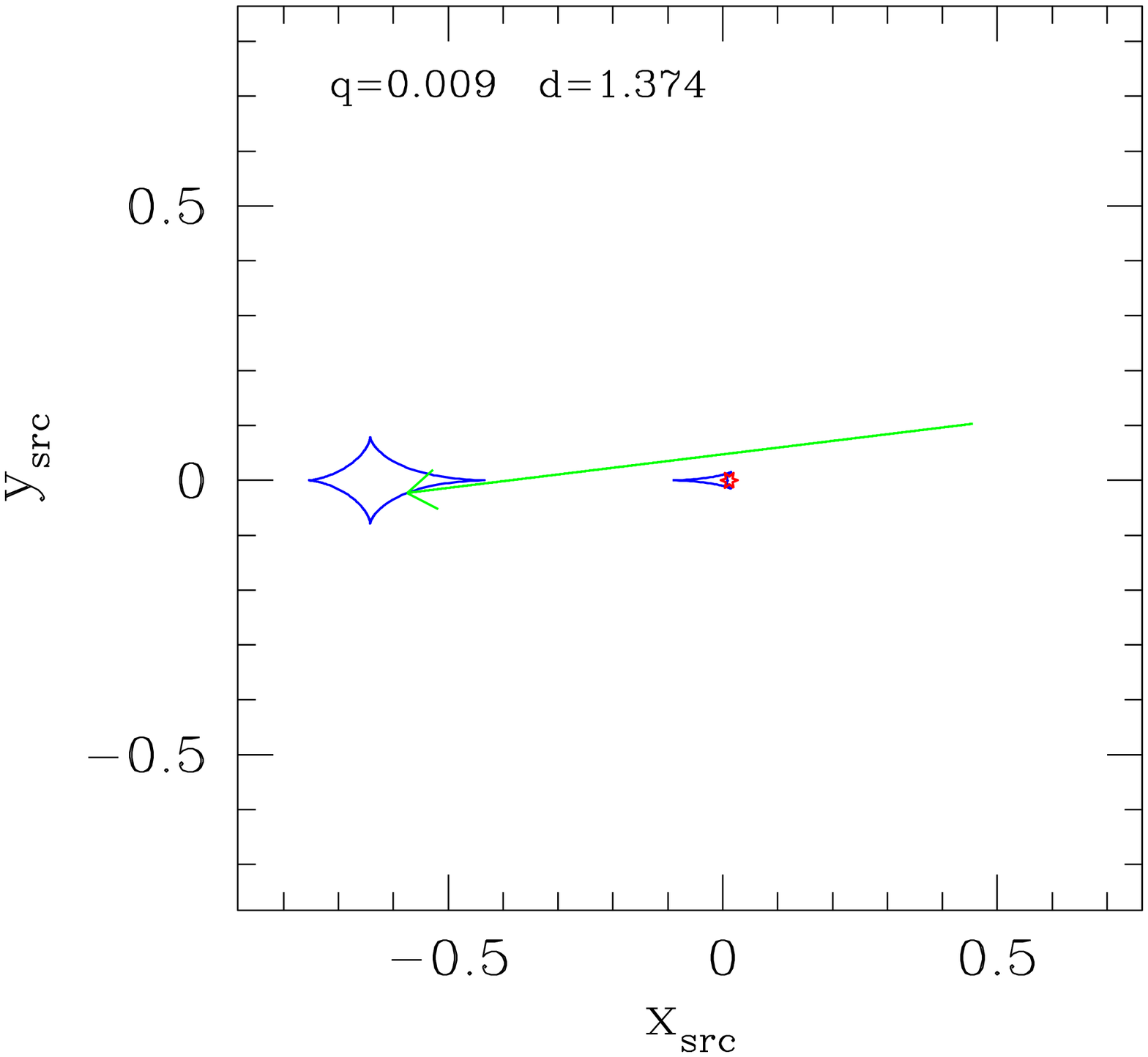}%
 \includegraphics[height=62mm,width=63mm]{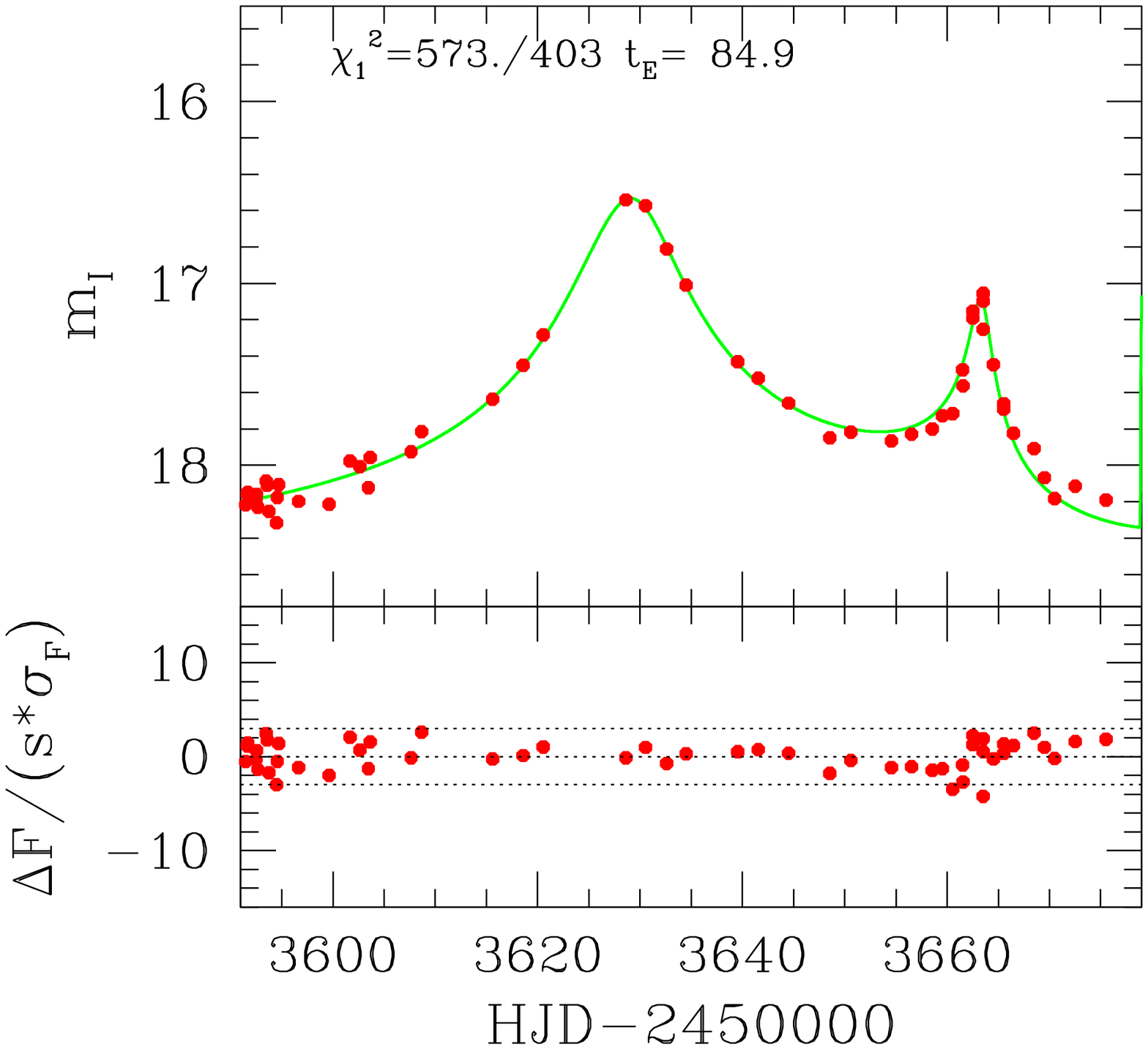}%

}

{\it Comparison of Binary Lens and Double Source Models}
\vskip6pt
Below we show the binary lens (on the left) and double source (on the
right) models of the light curves for some of the considered events. For
all binary lens models we have also calculated double source models. We do
not show events with evident jumps in the light curves, which can only be
modeled as caustic crossings. We include, however, a few cases with almost
smooth observed light curves, despite the fact that their binary lens
models are formally far better. The light curve in a double source model is
a sum of the constant blended flux plus the two single lens light curves
for the source components, each shown with dotted lines. In the plots we
use observed fluxes (not magnitudes) since they are additive, which is
important in double source modeling. In the majority of cases the binary
lens models give formally better fits as compared to the double source
models presented. On the other hand double source models, always producing
simpler light curves, look more natural in some cases.

\vskip 0.3cm

\noindent\parbox{12.75cm}{
\leftline {\bf OGLE 2005-BLG-017} 

\includegraphics[height=63mm,width=62mm]{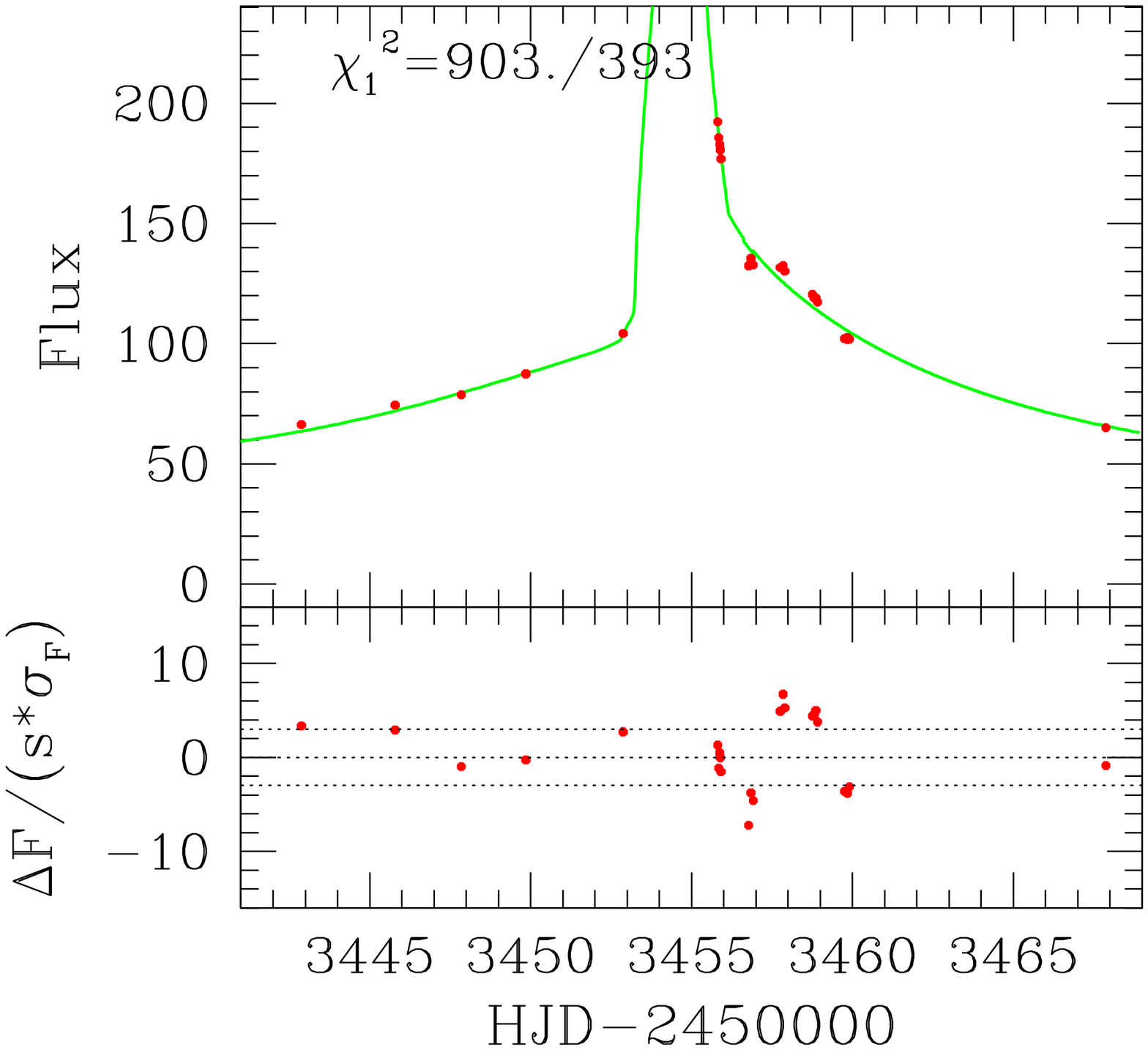}%
\includegraphics[height=63mm,width=62mm]{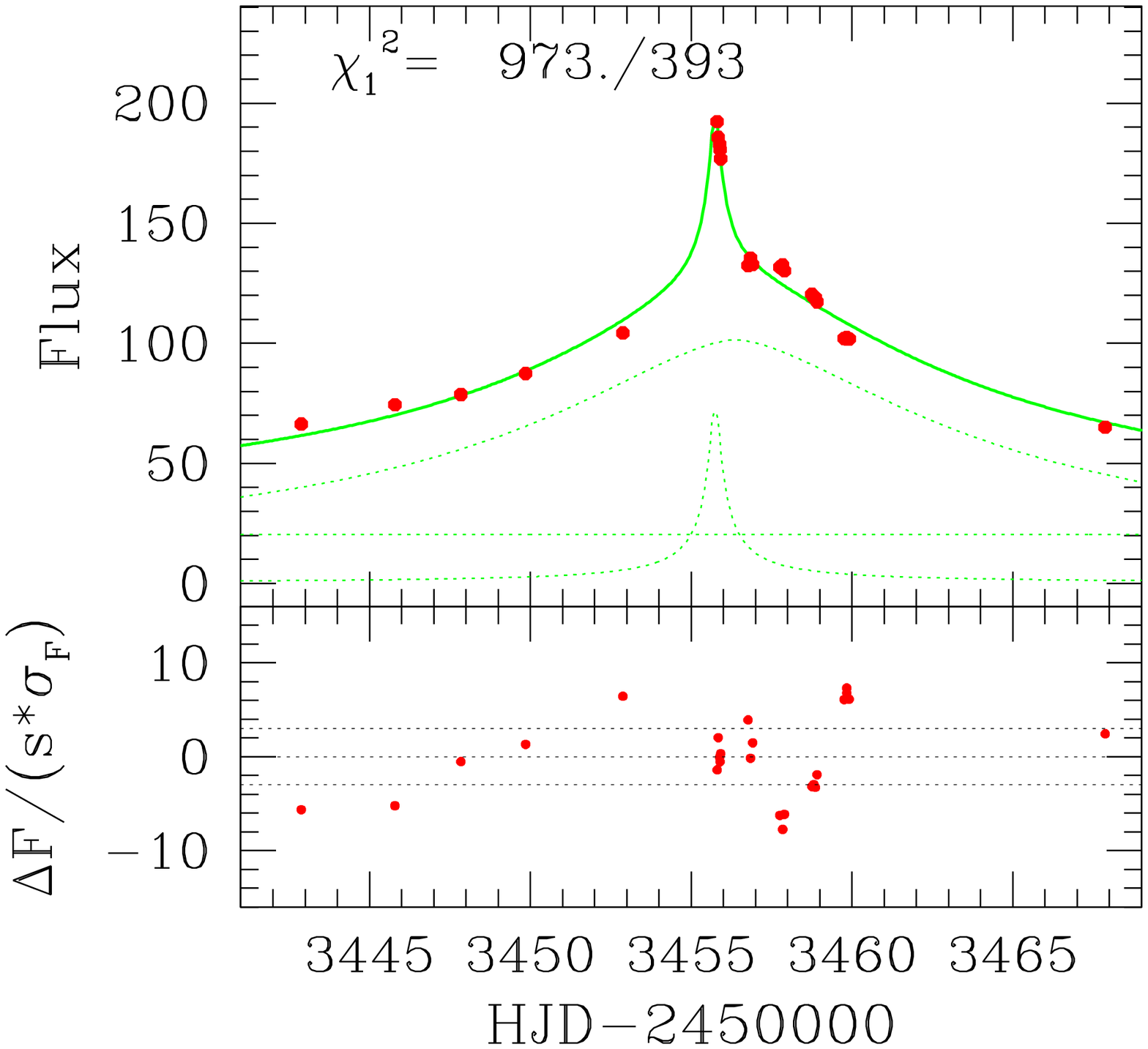}%

}

\noindent\parbox{12.75cm}{
\leftline {\bf OGLE 2005-BLG-018} 

 \includegraphics[height=63mm,width=62mm]{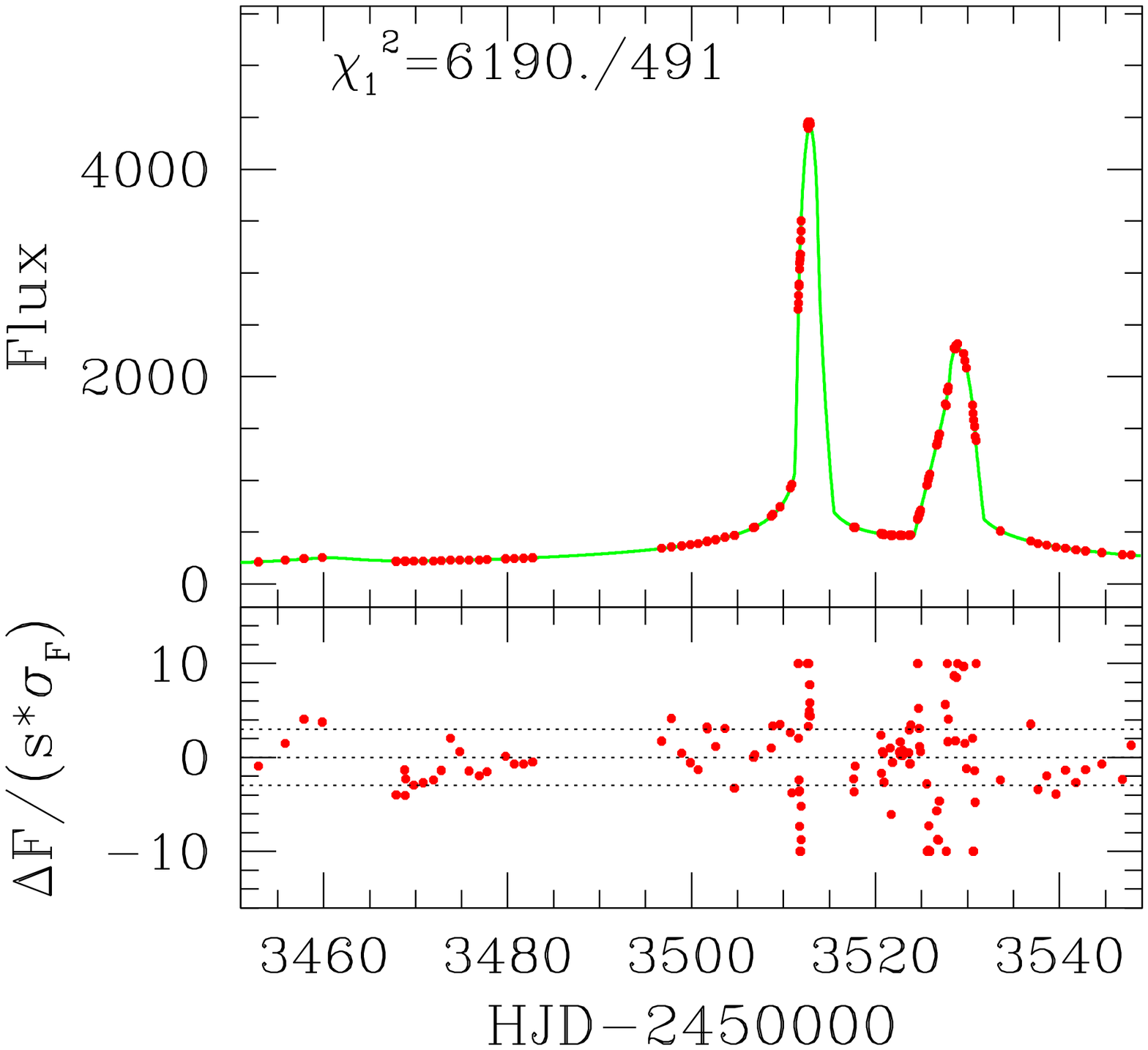}%
 \includegraphics[height=63mm,width=62mm]{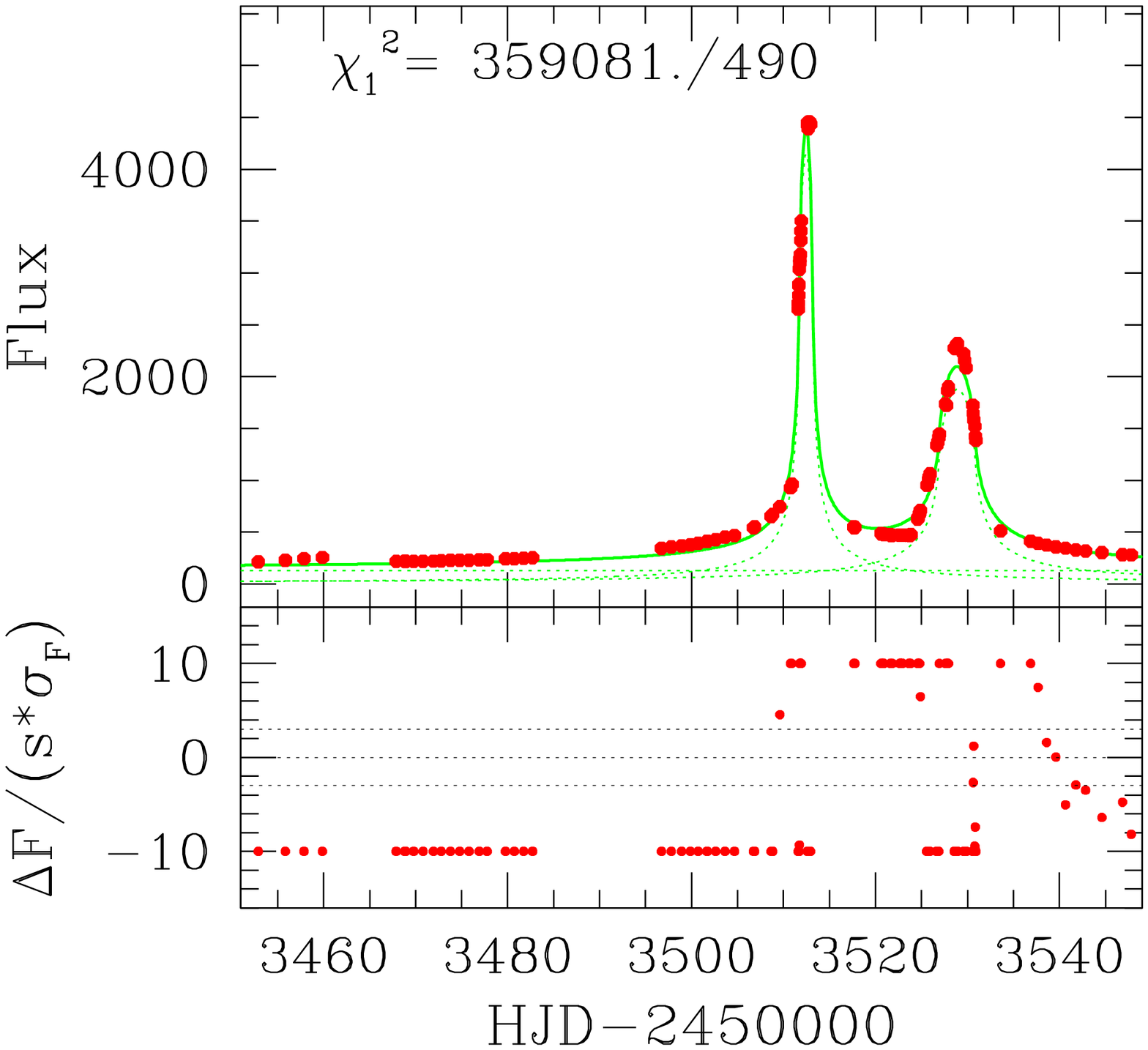}%

}

\noindent\parbox{12.75cm}{
\leftline {\bf OGLE 2005-BLG-062} 

 \includegraphics[height=63mm,width=62mm]{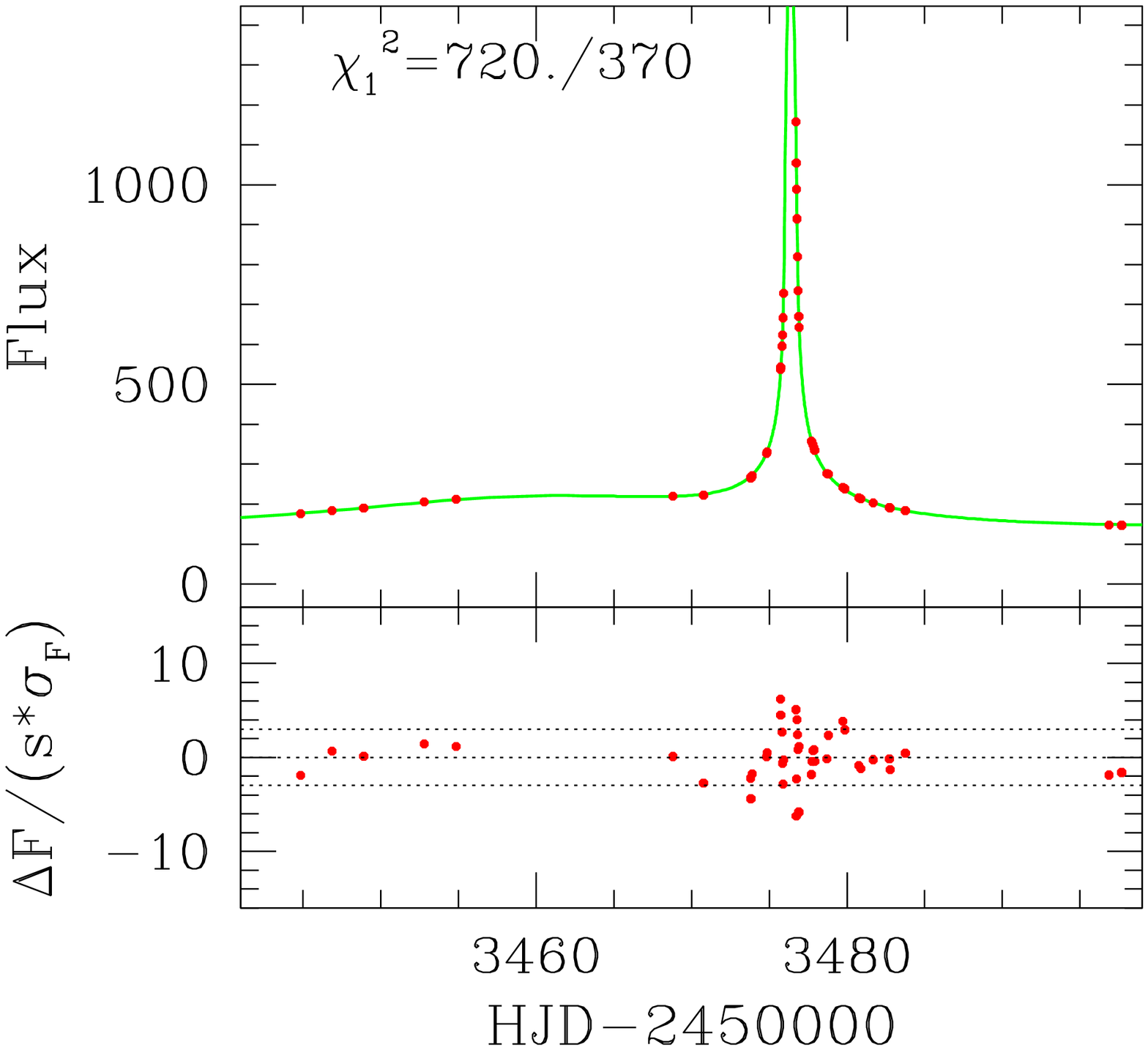}%
 \includegraphics[height=63mm,width=62mm]{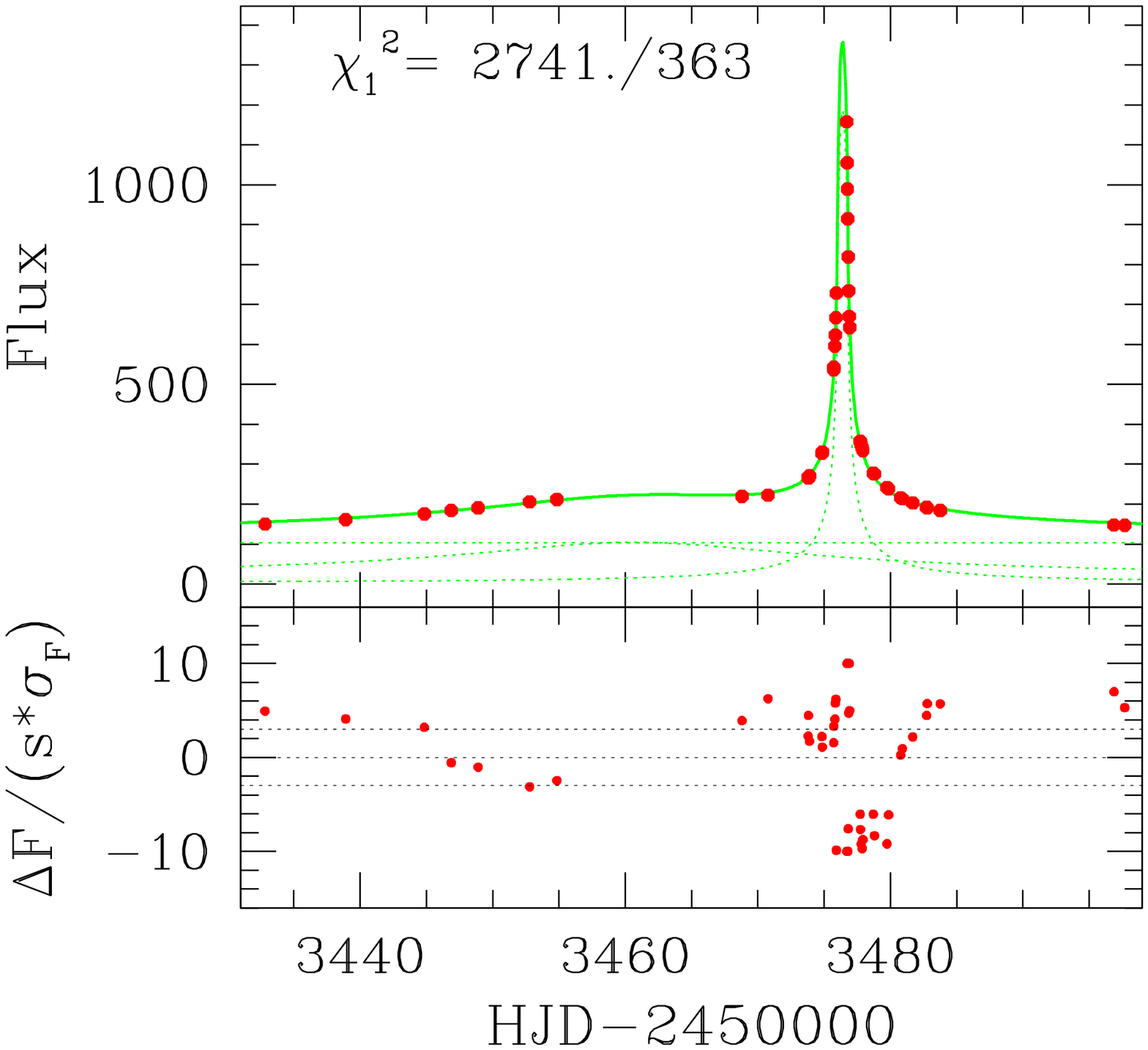}%

}

\noindent\parbox{12.75cm}{
\leftline {\bf OGLE 2005-BLG-153} 

 \includegraphics[height=63mm,width=62mm]{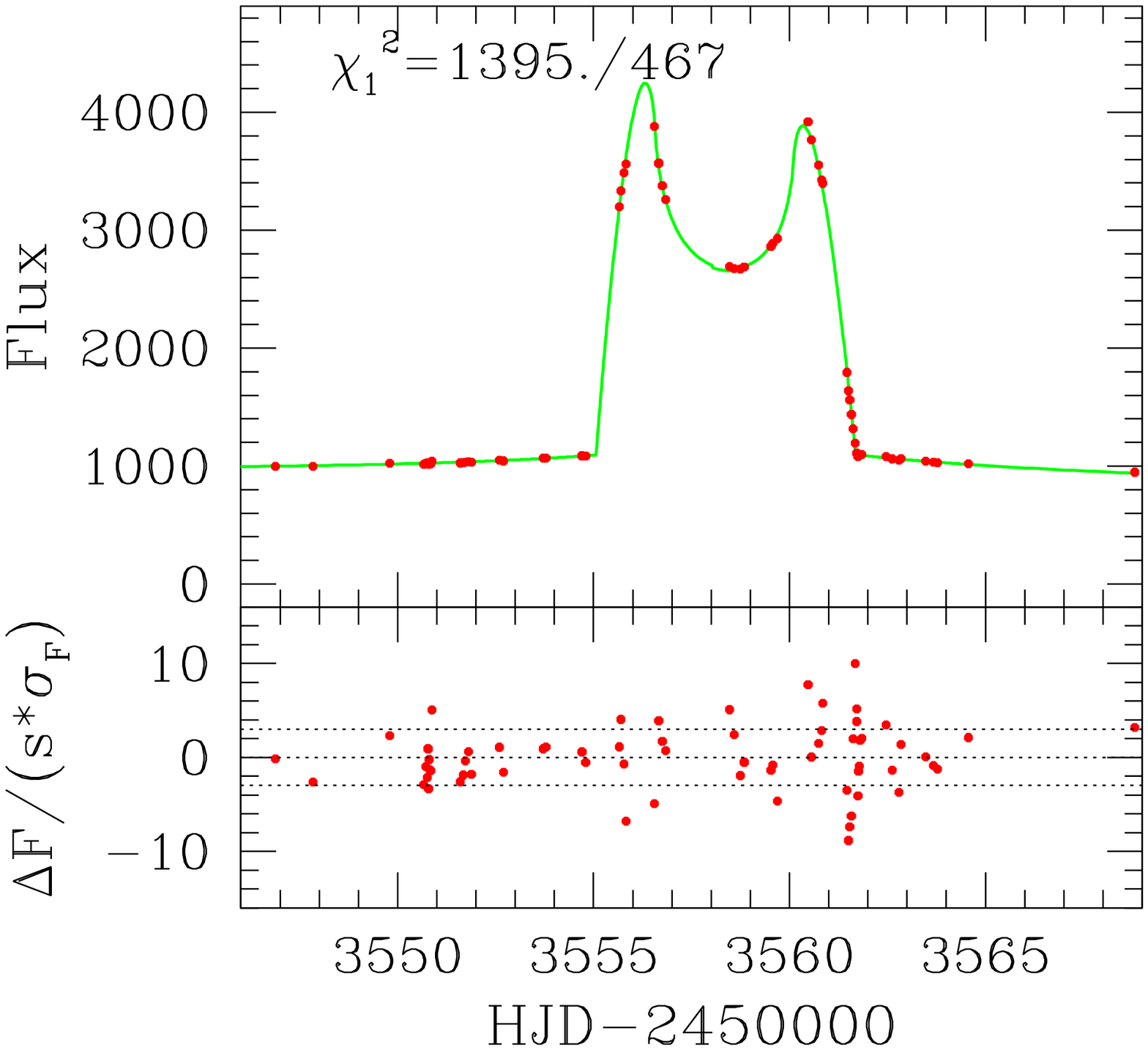}%
 \includegraphics[height=63mm,width=62mm]{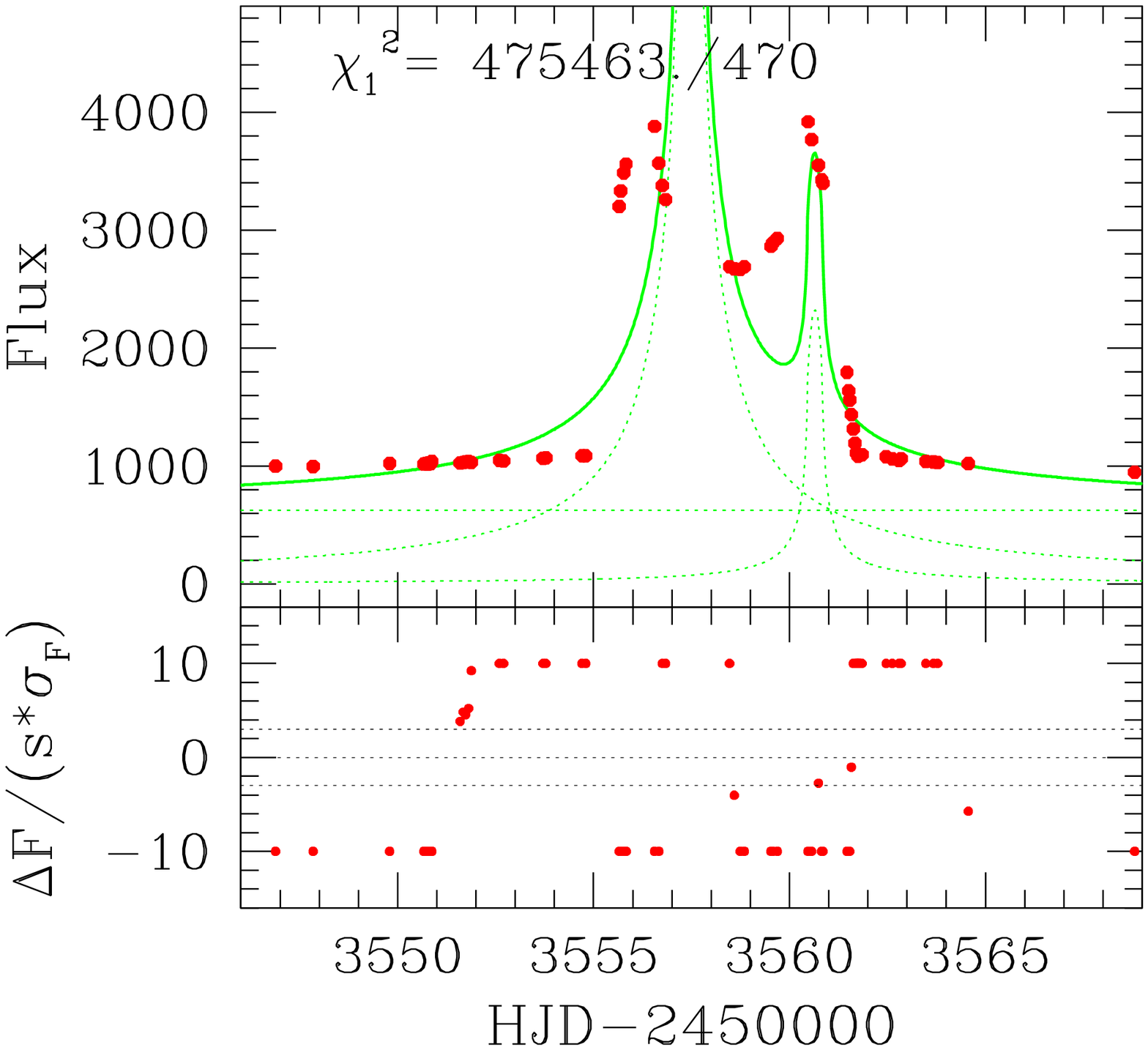}%

}

\noindent\parbox{12.75cm}{
\leftline {\bf OGLE 2005-BLG-226} 

 \includegraphics[height=63mm,width=62mm]{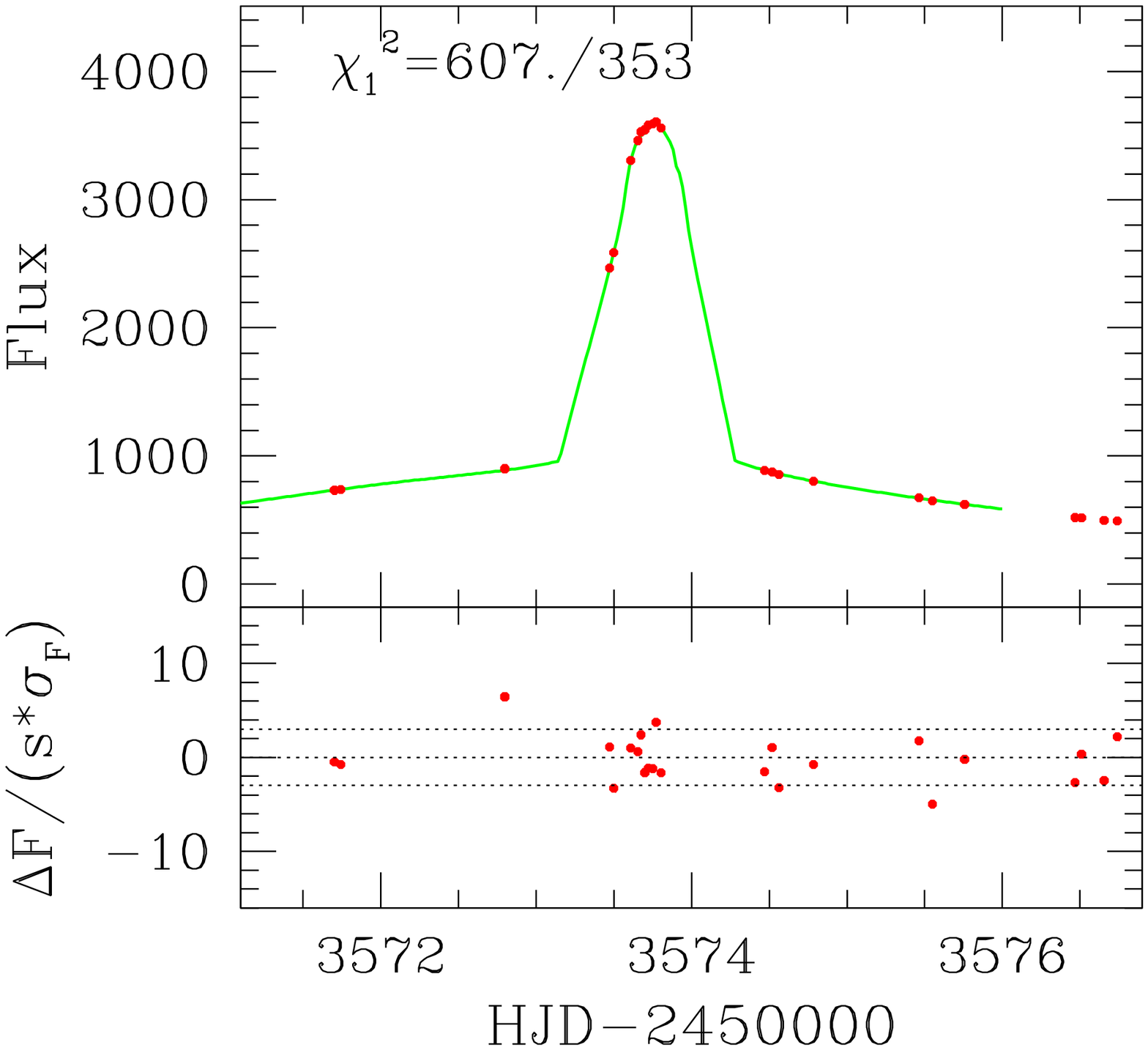}%
 \includegraphics[height=63mm,width=62mm]{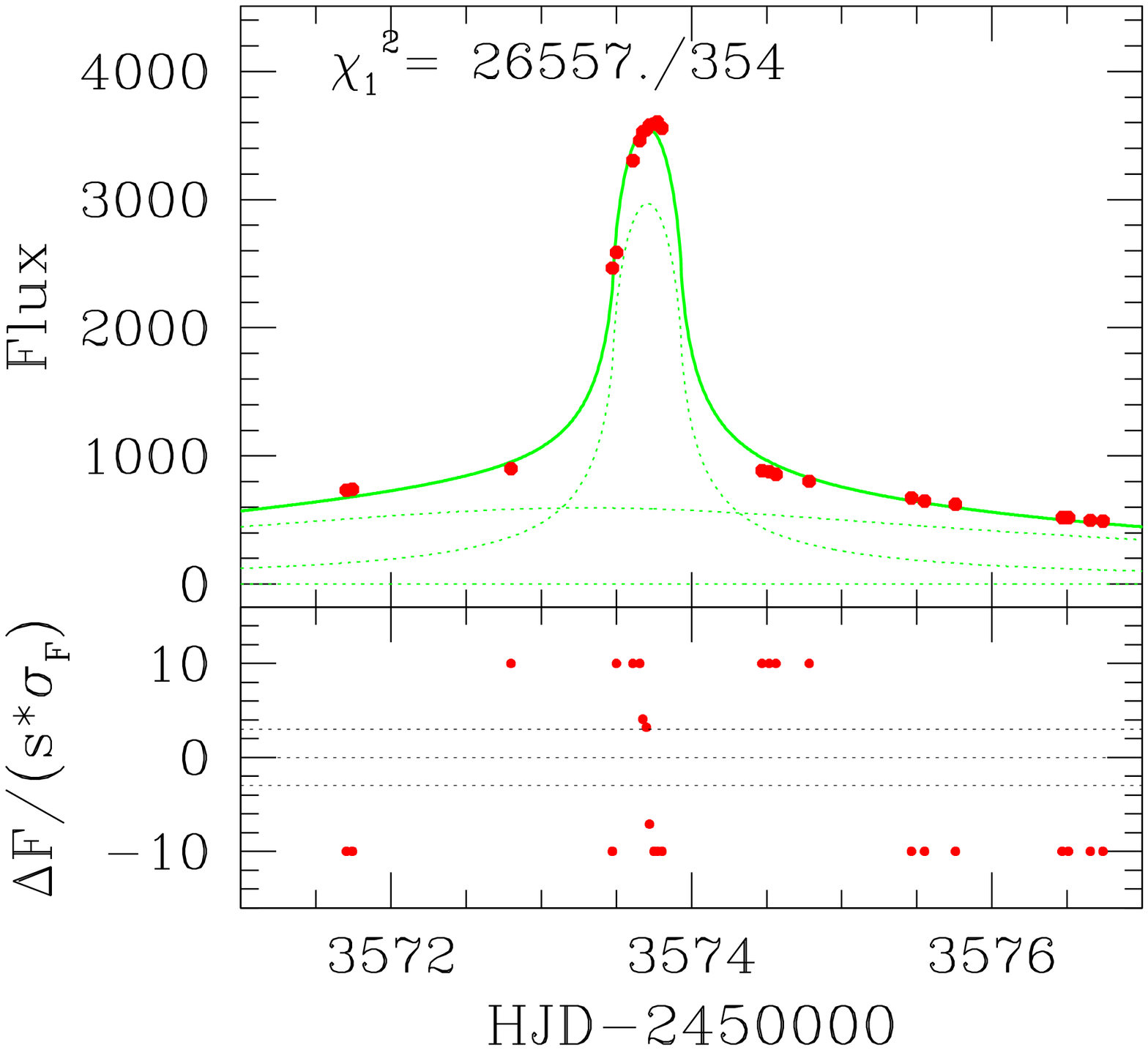}%

}

\noindent\parbox{12.75cm}{
\leftline {\bf OGLE 2005-BLG-477} 

 \includegraphics[height=63mm,width=62mm]{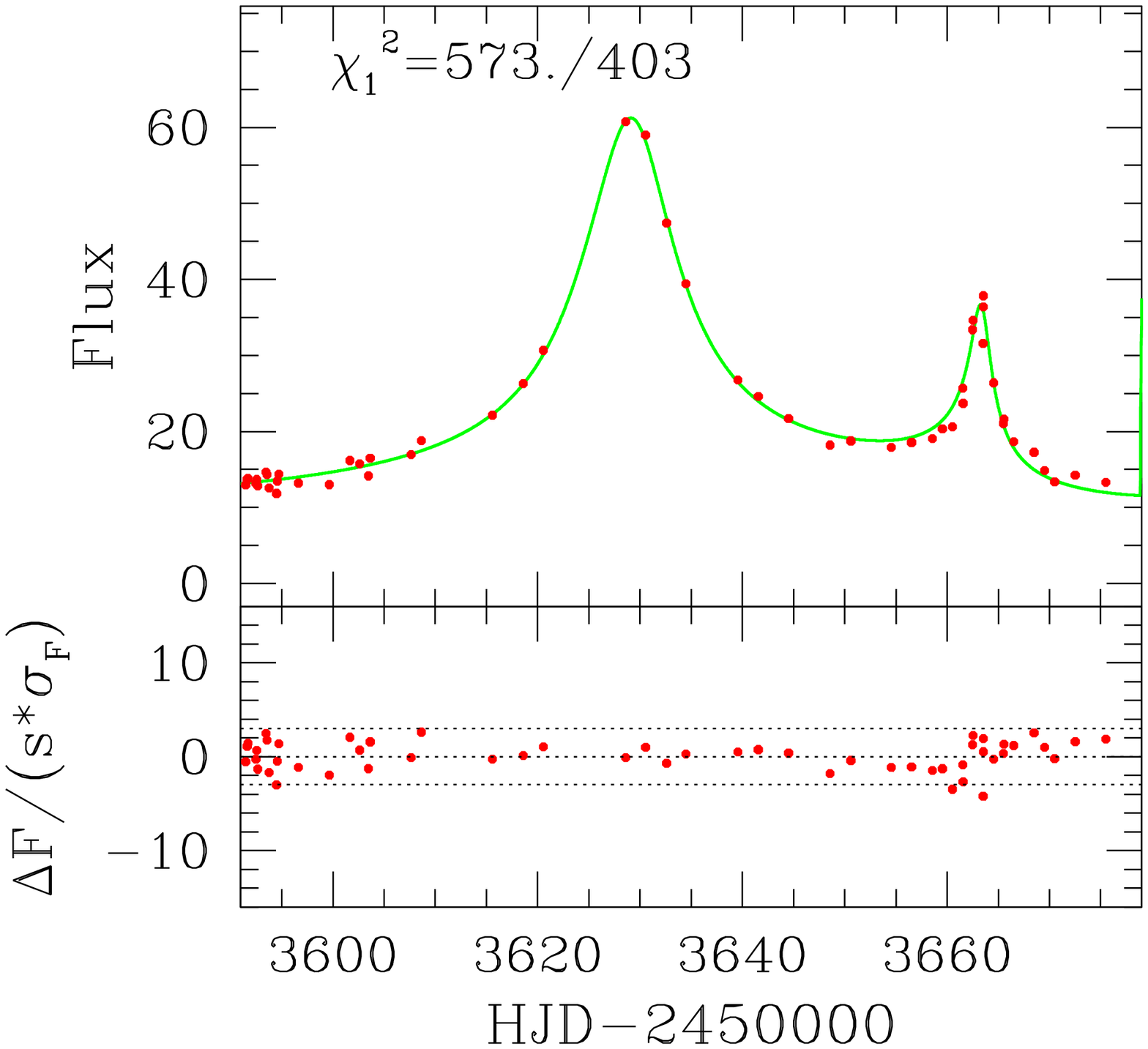}%
 \includegraphics[height=63mm,width=62mm]{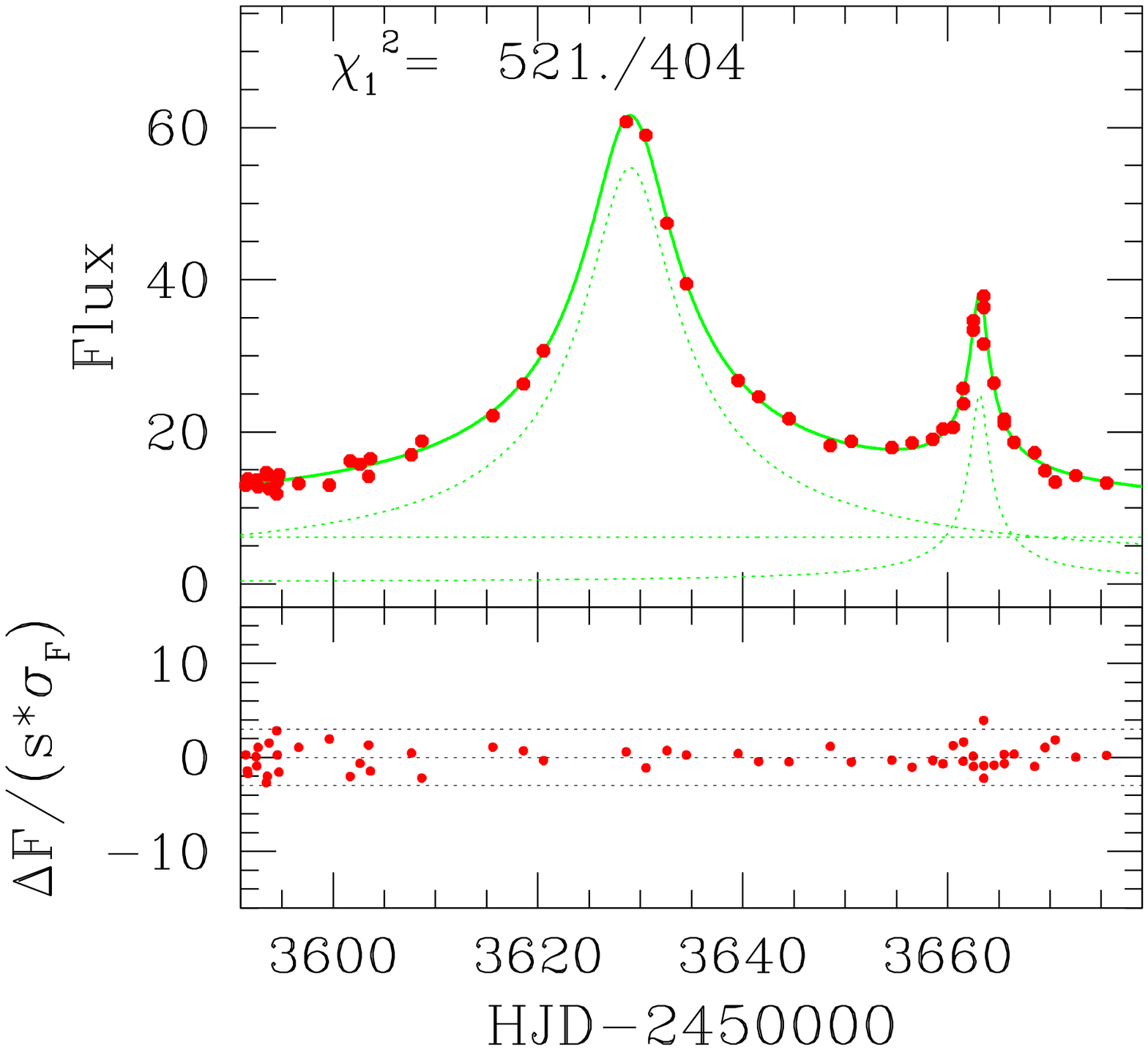}%

}
 
{\it Double Source Events}
\vskip6pt
\noindent\parbox{12.75cm}{
{\bf OGLE 2005-BLG-066}\hfill {\bf OGLE 2005-BLG-192} 

 \includegraphics[height=63mm,width=62mm]{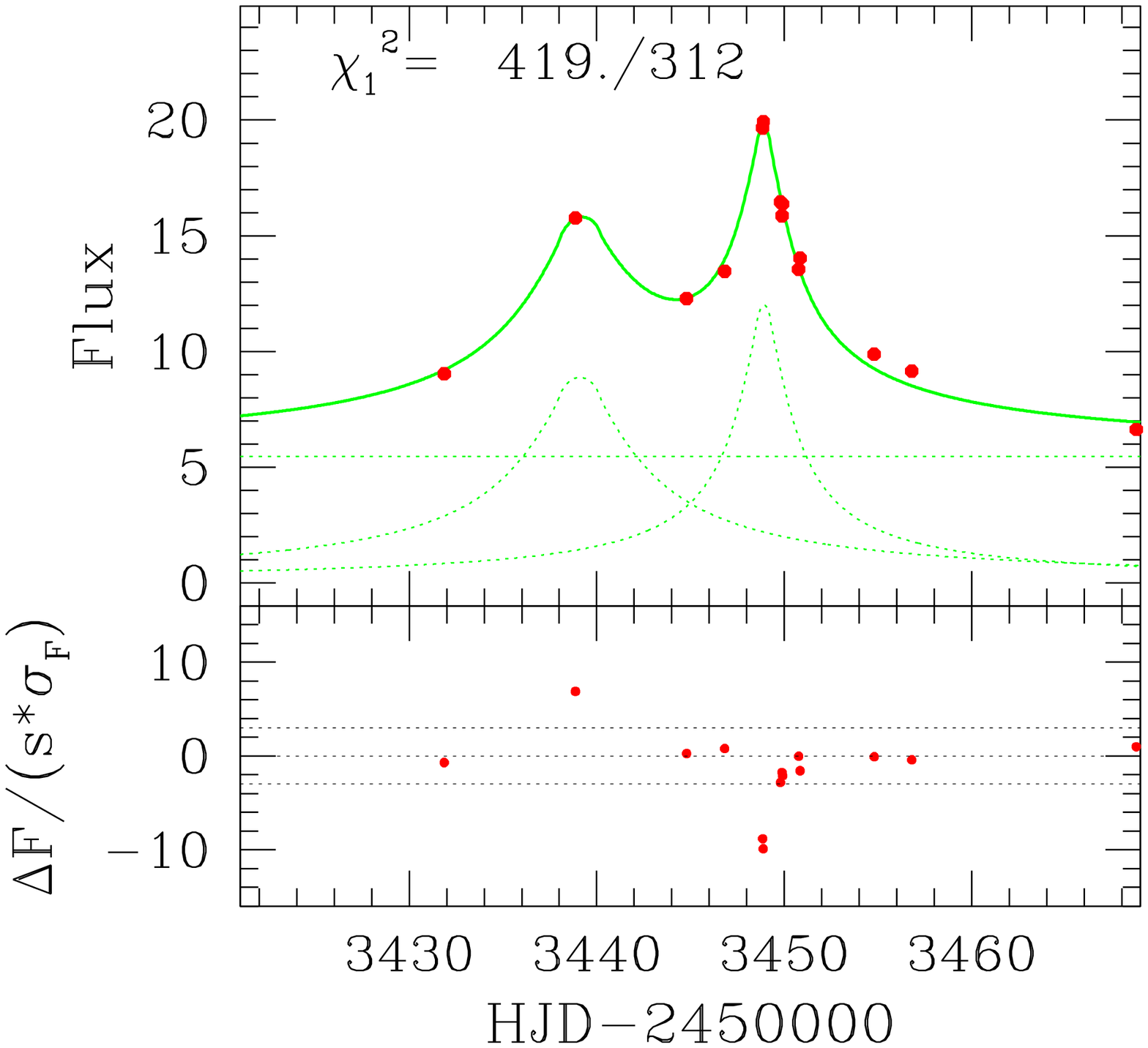}%
 \includegraphics[height=63mm,width=62mm]{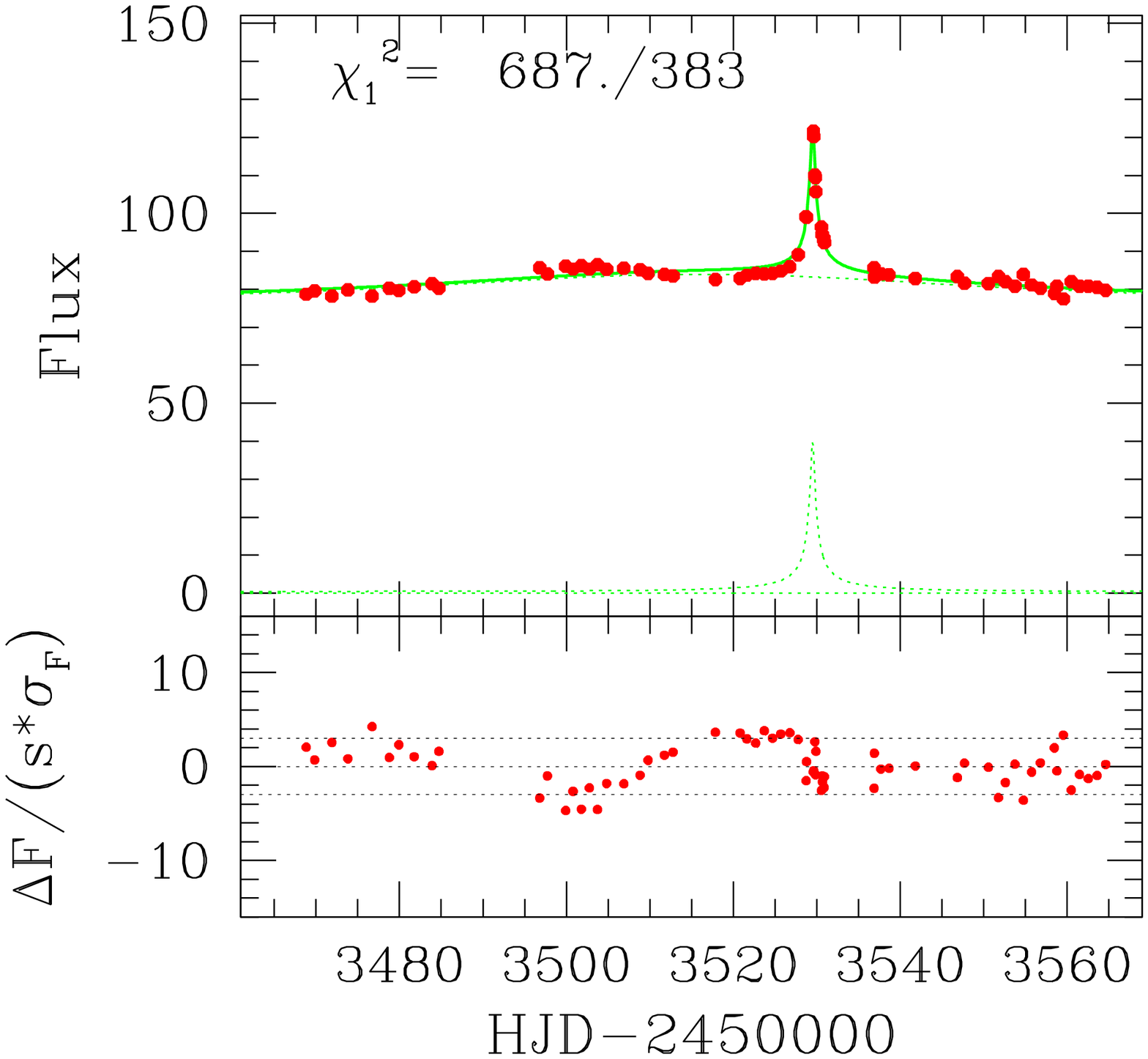}%

}
\newpage
{\it Unsuccessful Fit}
\vskip6pt
\noindent\parbox{12.75cm}{
{\bf OGLE 2005-BLG-331}\hfill  

 \includegraphics[height=63mm,width=62mm]{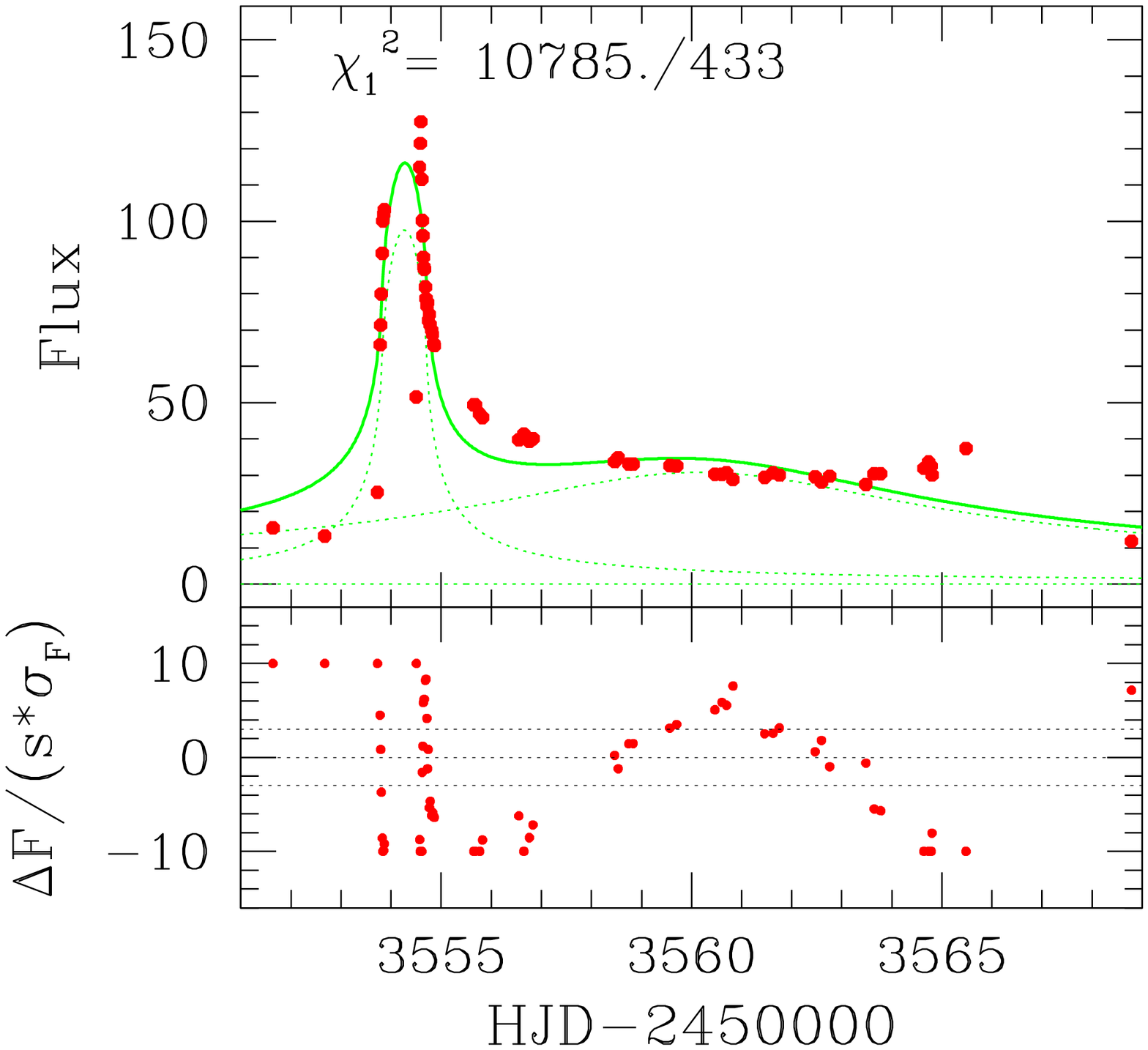} \hfill

}


\begin{references}
\refitem{Alard, C.}{2000}{\AAS}{144}{363}
\refitem{Alard, C., and Lupton, R.H.}{1998}{\ApJ}{503}{325}
\refitem{Albrow, M.D., \etal}{1999}{\ApJ}{522}{1022}
\refitem{Alcock, C., \etal}{2000}{\ApJ}{541}{270} 
\refitem{An, J.H., \etal}{2002}{\ApJ}{572}{521}
\refitem{Beaulieu, J.-P., \etal}{2006}{Nature}{439}{437}
\refitem{Bond, I.A., \etal}{2004}{\ApJL}{606}{L155} 
\refitem{DiStefano, R., and Mao, S.}{1996}{\ApJ}{457}{93}
\refitem{Dominik, M.}{1998}{\AA}{333}{L79}
\refitem{Dominik, M.}{2007}{\MNRAS}{377}{1679}
\refitem{Gaudi, B.S. and Han, Ch.}{2004}{\ApJ}{611}{528}
\refitem{Gould, A., \etal}{2006}{\ApJL}{644}{L37}
\refitem{Jaroszy\'nski, M.}{2002}{\Acta}{52}{39 (Paper I)}
\refitem{Jaroszy\'nski, M., and Paczy\'nski,B}{2002}{\Acta}{52}{361}
\refitem{Jaroszy\'nski, M., Udalski, A., Kubiak, M., Szyma\'nski, M.,
    Pietrzy\'nski, G., Soszy\'nski, I., \.Zebru\'n, K., Szewczyk, O., and
    Wyrzykowski, {\L}.}{2004}{\Acta}{54}{103 (Paper II)}
\refitem{Jaroszy\'nski, M., Udalski, A., Kubiak, M., Szyma\'nski, M.,
    Pietrzy\'nski, G., Soszy\'nski, I., \.Zebru\'n, K., Szewczyk, O., and
    Wyrzykowski, {\L}.}{2005}{\Acta}{55}{159}
\refitem{Jaroszy\'nski, M., Skowron, J., Udalski, A., Kubiak, M., 
    Szyma\'nski, M., Pietrzy\'nski, G., Soszy\'nski,~I., \.Zebru\'n, K., 
    Szewczyk, O., and Wyrzykowski, {\L}.}{2006}{\Acta}{56}{307 (Paper III)}
\refitem{Mao, S., and DiStefano, R.}{1995}{\ApJ}{440}{22}
\refitem{Mao, S., and Loeb, A.}{2001}{\ApJL}{547}{L97}
\refitem{Night, Ch., DiStefano, R., and Schwamb, M.}{2007}{\ApJ}{~}{submitted
(astro-ph/0705.0169)}
\refitem{Paczy\'nski, B.}{1996}{Ann. Rev. Astron. Astrophys.}{34}{419}
\refitem{Schneider, P., and Weiss, A.}{1986}{\AA}{164}{237}
\refitem{Udalski, A.}{2003}{\Acta}{53}{291}
\refitem{Udalski, A., \etal}{2005}{\ApJL}{628}{L109}
\refitem{Udalski, A., Szyma\'nski, M., Kaluzny, J., Kubiak, M., Mateo, M., 
Krzemi\'nski, W., and Paczy\'nski, B.}{1994}{\Acta}{44}{227} 
\refitem{Wyrzykowski, {\L}.}{2005}{\it PhD Thesis}{}{Warsaw University
Astronomical Observatory}
\end{references}
\end{document}